\documentclass[useAMS,usenatbib]{mn2e}

\usepackage{times}  
\usepackage{listings}
\usepackage{graphics}
\usepackage{subfigure}
\usepackage{graphicx}
\usepackage{amssymb}
\usepackage{hyperref}
\usepackage{aas_macros}

\def\simprop{ \lower .75ex \hbox{$\sim$} \llap{\raise .27ex \hbox{$\propto$}} }

\title[II. Misaligned gas accretion]{
The origin of the atomic and molecular gas contents of early-type galaxies. II. Misaligned gas accretion}
\author[Claudia del P. Lagos et al.]{
\parbox[t]{\textwidth}{
\vspace{-1.0cm}
Claudia del P. Lagos$^{1}$,
Nelson D. Padilla$^{2,3}$, 
Timothy A. Davis$^{1,4}$, 
Cedric G. Lacey$^{5}$, 
Carlton M. Baugh$^{5}$,
Violeta Gonzalez-Perez$^{5}$,
Martin A. Zwaan$^{1}$,
Sergio Contreras$^{2}$
}
\vspace*{6pt} \\
$^{1}$European Southern Observatory, Karl-Schwarzschild-Strasse 2, 85748, Garching, Germany.\\
$^{2}$Instituto de Astrof\'isica, Pontificia Universidad
Cat\'olica de Chile, Av. Vicu\~na Mackenna 4860, Santiago., Chile.\\
$^{3}$Centro de Astro-Ingenier\'ia, Pontificia Universidad Católica de Chile, Av. Vicu\~na Mackenna 4860, Santiago, Chile.\\
$^{4}$Centre for Astrophysics Research, Science \& Technology Research Institute, University of Hertfordshire, Hatfield, AL10 9AB, UK.\\
$^{5}$Institute for Computational Cosmology, Department of Physics, University of Durham, South Road, Durham, DH1 3LE, UK.
\vspace*{-0.5cm}}

\begin{document}


\pagerange{\pageref{firstpage}--\pageref{lastpage}} \pubyear{2011}

\maketitle

\label{firstpage}

\begin{abstract}
We study the origin of the wide distribution of 
angles between the angular momenta of the stellar and gas components, $\alpha_{\rm G,S}$, in early-type galaxies (ETGs). 
We use the {\texttt{GALFORM}} model of galaxy formation, set
in the $\Lambda$ cold dark matter framework, and coupled it with a 
Monte-Carlo simulation to follow the angular momenta flips driven by matter accretion onto haloes and galaxies. 
We consider a gas disk to be misaligned with respect to the stellar body if $\alpha_{\rm G,S}>30$~degrees. 
By assuming that the only sources of misaligments in galaxies are galaxy mergers, we place 
a lower limit of $2-5$ per cent on the fraction of ETGs with misaligned gas/stellar components. 
These low fractions are inconsistent with the observed value of  
$\approx 42\pm 6$ per cent in ATLAS$^{\rm 3D}$.
In the more general case, in which smooth gas accretion in addition to galaxy mergers can drive misalignments, our calculation 
 predicts that $\approx 46$ per cent  of ETGs have
$\alpha_{\rm G,S}>30$~degrees.
In this calculation, we find correlations between $\alpha_{\rm G,S}$ 
and stellar mass, cold gas fraction and star formation rate, such that 
 ETGs with high masses, low cold gas fractions and low star formation rates are more likely to display 
aligned cold gas and stellar components. We confirm these trends observationally for the first time using ATLAS$^{\rm 3D}$ data.
We argue that the high fraction
of misaligned gas discs observed in ETGs is mostly due to smooth gas accretion (e.g. cooling from the hot halo 
of galaxies) which takes place after most of the stellar mass 
of the galaxy is in place and comes misaligned with respect to the stellar component. 
Galaxies that have accreted most of their cold gas content prior to the time where most of the stellar mass was in place 
show aligned components.
\end{abstract}


\section{Introduction}

Contrary to naive expectations, early-type galaxies (ETGs; ellipticals and lenticulars), 
 host significant quantities of cold gas in the form of atomic hydrogen (HI) and 
molecular hydrogen (H$_2$), { although the gas fractions in ETGs are on average lower than in late-type galaxies} 
(e.g. \citealt{Gallagher75}; \citealt{Wardle86}; \citealt{Wiklind89}; 
\citealt{Morganti06}; \citealt{Oosterloo07}; \citealt{Welch10}; \citealt{Young11}; \citealt{Serra12}). 
\citet{Serra12} found that $42$ per cent of ETGs in the ATLAS$^{\rm 3D}$ 
have HI masses $M_{\rm HI}>10^7\,M_{\odot}$, while \citet{Young11} show 
that $23$ per cent of ETGs have $M_{\rm H_2}>10^7\,M_{\odot}$ in the same survey. Most ETGs with detected HI and/or H$_2$ 
show settled configurations, with the cold gas being in a disk or a ring (\citealt{Serra12}; \citealt{Davis13}). 
Only $30$ per cent of ETGs with 
detected HI show unsettled configurations. This has been interpreted as being due to interactions with the 
group environments or a recent minor galaxy merger \citep{Serra12}. This percentage decreases when focusing on 
H$_2$ only (\citealt{Davis13}; \citealt{Alatalo13}). 

\citet{Lagos14} (hereafter Paper I) 
explored the origin of the HI and H$_2$ gas contents of ETGs using a semi-analytic model of galaxy formation and showed 
that the observed gas fractions in ETGs arise in the model 
due to inefficient replenishment of the gas in ETGs as a result of both 
ram pressure stripping of the hot gas and heating by active galactic nuclei (AGN), which prevents gas cooling 
from the hot halo. Model ETGs with high HI and H$_2$ gas contents were shown to be hosted by low mass haloes, 
and have, on average, `younger morphologies' (i.e. smaller look-back times to the last time these ETGs showed late-type 
morphologies). These ETGs were shown in Paper I to be more isolated, which agrees  
with observational inferences of lower density environments for the gas-rich ETGs \citep{Young13}.

An important finding in the exhaustive observations of the kinematics of the different mass components of ETGs 
was presented in \citet{Davis11}, where it was convincingly shown that $\approx 42$ per cent of ETGs have 
ionised gas misaligned with the stellar kinematics (i.e. the angle between the angular momenta 
of the stars and the gas is $>30$~degrees). 
The way the angle between the different galaxy components is measured in the observations is using 
high signal-to-noise ($S/N>40$ ; \citealt{Emsellem04}) integral field spectroscopy and millimeter wave interferometry, 
from which one can construct two-dimensional kinematic maps.
\citet{Davis11} calculated position angles from the 2D kinematic maps of the stellar and gas components in a way that 
they trace the bulk of the components rather than substructures. From the position angles, 
 the projected misalignments, rather than the three-dimensional misalignments, were measured.
Typical uncertainties in the measured position angles in ATLAS$^{\rm 3D}$ are
 of $10$~degrees. \citet{Davis11} also show that the molecular and ionised gas components 
are aligned with each other, suggesting a common origin. 

The high fraction of stellar/gas misalignments reported by \citet{Davis11} led to  
 speculation that the origin of this gas is external, pointing to minor mergers as the dominant source 
of the gas in ETGs that show misaligned gas disks (e.g. \citealt{Davis11}; \citealt{Serra12}). 
This intriguing inference has been explored in simulations of galaxy formation very recently, with 
small samples of simulated galaxies. For example, \citet{Serra14} show that 
  although simulations can reproduce the nature of slow and fast rotators
of the early-type population (see also \citealt{Naab13}), 
the HI contents predicted by these simulations are too low, while also being almost 
always kinematically aligned with the stellar component. The main disadvantage of the work of Serra et al. is that only $50$ 
simulated galaxies were analysed and therefore strong conclusions regarding the consistency of the simulations with 
observations cannot be reached.

An important caveat in the interpretation presented by \citet{Davis11} and \citet{Serra12}, 
in which only minor galaxy mergers account for the misaligned gas disks observed in ETGs, is that it ignores the 
stochastic nature of matter accretion from sources (other than mergers) 
predicted in a $\Lambda$ cold dark matter (CDM) universe 
(e.g. \citealt{Dekel09}; \citealt{Johansson12}).
It implies that gas accretion from sources other than galaxy mergers, 
such as cooling from the hot halo and collimated inflows of gas from filaments, is always aligned with the stellar component. 
\citet{Sales12}, \citet{Sharma12}, \citet{Bett12}, \citet{Aumer13} and \citet{Padilla14} show 
strong evidence from simulations that in a $\Lambda$CDM universe the angular momentum of galaxies is not necessarily aligned with that of 
their host haloes, as accretion from either the cosmic web or mergers can be stochastic.
\citet{Bett12} go a step further and show that the inner parts
of haloes, where galaxies live, can suffer much more frequent flips of their angular momentum vector than the halo as a whole 
(see also \citealt{Sharma12}).
 
\citet{Bett10} show that the median angle between the inner ($\lesssim 0.25\,R_{\rm vir}$)
and total ($\le R_{\rm vir}$) angular momentum vectors is $\approx 25$~degrees. Bett et al. also show that the inclusion of baryons
drives even larger misalignments, with half of the galaxies having their spin axis misaligned by more than $45$~degrees
from the host halo spin (see also \citealt{Bryan13} and \citealt{Tenneti14}). 
There is also evidence from hydro-dynamical simulations that the angular momentum of galaxies 
is affected by the large scale structure; i.e. filamentary structure can fuel gas into galaxies changing its angular momentum 
direction (e.g. \citealt{Danovich14}; \citealt{Dubois14}). The result of such accretion process can be alignment with respect to 
large structures, outside haloes, particularly in the low mass regime (\citealt{Dubois14}). 

Our motivation is therefore to investigate the alignments between the galaxy components, stars and cold gas (atomic and 
molecular gas), and the dark matter halo 
by following the flips in the angular momentum of ETGs throughout their growth history. 
{By flips we mean any change in the direction of the angular momentum vector}.
This allows us to statistically assess 
the probability of having a present day ETG with a gas disk that is misaligned with respect to the stellar component. 
For consistency with the measured projected misalignment angle between different galaxy components in the observations, we show throughout the paper 
predictions of projected misalignments.
The study of angular momentum in galaxies is particularly relevant as ongoing and future surveys, such as 
Mapping Nearby Galaxies at APO (MaNGA\footnote{\tt https://www.sdss3.org/future/manga.php}), 
Calar Alto Legacy Integral Field spectroscopy Area survey (CALIFA\footnote{{\tt http://califa.caha.es/} \citep{Husemann13}.}), 
 the Sydney-Australian-Astronomical-Observatory Multi-object Integral-Field Spectrograph 
(SAMI\footnote{{\tt http://sami-survey.org/} \citep{Croom12}.}),
and ultimately the Square Kilometer Array (SKA\footnote{\tt https://www.skatelescope.org/}), 
 promise to transform our understanding of angular momentum 
in galaxies in their different components. An example of this is the mass-spin-morphology relation
presented by \citet{Obreschkow14b}.

For this study we use the semi-analytical model {\texttt{GALFORM}}
in a $\Lambda$CDM cosmology (\citealt{Cole00}) presented by Lacey et al. (2014, in prep.).
This model includes the improved treatment of star formation (SF) implemented by \citet{Lagos10,Lagos11}. This extension
 explicitly splits the hydrogen content of the ISM
of galaxies into HI and H$_2$. In Paper I we show that the Lacey et al. model provides a very good representation 
of the gas contents of ETGs (in the form of HI and H$_2$) in the local Universe, particularly when gradual ram pressure 
stripping of the hot gas is included (see \citealt{Lagos14b} for predictions of the gas content of galaxies in the Lacey et al. model 
at high-redshifts). 

This paper is organised as follows. In $\S 2$ we summarise the main aspects of the 
{\tt GALFORM} model and the flavour presented by Lacey et al. (2014).
In $\S 3$ we summarise the method for following the flips in the angular momenta of the different component of galaxies 
(stars, cold gas and dark matter halo), originally introduced by \citet{Padilla14}.
In $\S 4$, we present the growth history of ETGs and analyse the transient nature of galaxy morphologies.
In $\S 5$, we describe how we calculate the angular momentum flips throughout the history of ETGs and present the 
expectation for the number of misaligned gas disks and compare with observations. We also introduce 
a new scenario that can lead to gas disks becoming misaligned in addition to galaxy mergers. 
In $\S 6$ we discuss limitations of the model presented here, also showing predictions for late-type galaxies, 
and give our conclusions in $\S 7$.

\section{Modelling the morphological evolution, neutral gas content and star formation in galaxies}\label{modelssec}

In paper I, we provided a detailed description of all the relevant physical mechanisms which affect the history of 
ETGs, such as disk formation, bulge formation, ram pressure stripping of the hot gas, star formation and recycling 
of mass and metals in stars. Here we outline the processes that are modelled in {\texttt{GALFORM}} 
and give a brief overview of the model we adopt as a standard in this paper to study the alignment between 
cold gas and stars in galaxies, which is a variant of the model of Lacey et al. (2014).

{\tt GALFORM} 
accounts for the main physical processes
that shape the formation and evolution of galaxies. These are: (i) the collapse
and merging of dark matter (DM) haloes, (ii) the shock-heating and radiative cooling
of gas inside DM haloes, leading to the formation of galactic disks, (iii) star
formation in galaxy disks (quiescent mode), (iv) feedback
from supernovae (SNe), from heating by active galactic nuclei (AGN) 
 and from photo-ionization of the
inter-galactic medium (IGM), (v) chemical 
enrichment of stars and gas, (vi) galaxy mergers driven by
dynamical friction within common DM haloes which can trigger bursts of star formation,
and lead to the formation of spheroids, (vii) global disk instabilities, which also lead to the formation of spheroids, 
and (viii) gradual ram pressure stripping of the hot gas.
Galaxy luminosities are computed from the predicted star formation and
chemical enrichment histories using a stellar population synthesis model (see \citealt{Gonzalez-Perez13}).
{Note that in the literature
`quiescent' is generally used to refer to passive galaxies and/or to those with low  
star formation rates (SFRs) compared to the median at a given stellar mass, but here we use it to distinguish 
star formation taking place in the disk of galaxies from the starburst mode, 
which takes place exclusively in the central spheroid of the galaxy and at efficiencies that are generally higher.}

Here we focus on the variant of {\tt GALFORM} presented in Lacey et al. (2014; hereafter Lacey14) 
which includes all the processes listed above. One important feature of the Lacey14 model is that  
it adopts a non-universal stellar initial mass function (IMF). 
The IMF describing SF in disks (i.e. the quiescent mode) is the \citet{Kennicutt83} IMF\footnote{The distribution of the masses of stars
produced follows ${\rm d}N(m)/{\rm d\, ln}\,m \propto m^{-x}$, where $N$ is the number of stars of mass $m$ formed,
 and $x$ is the IMF slope. For a \citet{Kennicutt83} IMF, $x=1.5$ for masses in the range $1\,M_{\odot}\le m\le 100\,M_{\odot}$ and
$x=0.4$ for $m< 1\,M_{\odot}$.}, while 
 a more top-heavy IMF is adopted in starbursts (i.e. with an IMF slope $x=1$). This is inspired by \citet{Baugh05}, who used a top-heavy 
IMF to reconcile the model predictions with observations of
the number counts and redshift distribution of submillimeter galaxies.
 We note, however, that Baugh et al. adopted a more top-heavy IMF for starbursts, with
 $x=0$, than is used for Lacey14. 

We now give some details of each of the processes above as included in Lacey14:
\begin{itemize}
\item In Lacey14 the halo merger trees are extracted from the WMAP7 \citep{Komatsu11} version of the 
Millennium cosmological $N$-body simulation \citep{Springel05} (refer to as MS-W7 simulation). 
The cosmological parameters are  $\Omega_{\rm m}=\Omega_{\rm DM}+\Omega_{\rm baryons}=0.272$ (with a
baryon fraction of $0.167$), $\Omega_{\Lambda}=0.728$, $\sigma_{8}=0.81$
and $h=0.704$.
\item Gas cooling is calculated assuming that gas in haloes follows a $\beta$ profile \citep{Cavaliere76}. 
The amount of cooling then depends on the gas density and its metallicity following the 
tabulated cooling function of \citet{Sutherland93}. 
The amount of gas that is added to the disk depends on the cooling time and the free-fall time (see \citealt{Cole00} and \citealt{Benson10}). 
Given the time elapsed since the formation of the DM halo (i.e. the last mass doubling), the free-fall radius, $r_{\rm ff}$, 
is the maximum radius in the hot halo from which material could have moved to the disk; the cooling radius, $r_{\rm cool}$, 
encloses the radius within which gas has had time to cool.
The mass accreted onto the disk simply corresponds to the
hot gas mass enclosed within $r={\rm min}[r_{\rm cool},r_{\rm ff}]$.
\item Lacey14 adopts two different SF laws, one for quiescent SF (i.e. taking place in the disk) 
and another for starbursts (driven by galaxy mergers and disk instabilities). 
In the case of quiescent SF, Lacey14 adopts the SF law introduced by \citet{Blitz06}, where 
stars form from the molecular gas in the disk, and the partition of atomic and molecular gas 
depends on the hydrostatic pressure in the midplane of the disk (see \citealt{Lagos10} for details). 
For starbursts, the SFR is calculated from the available cold gas mass (HI plus H$_2$), regulated by a 
SF timescale, $\tau_{\rm SF}$, which depends on the dynamical timescale of the bulge, $\tau_{\rm dyn}$, 
$\tau_{\rm SF}=\rm max(\tau_{\rm min},f_{\rm dyn}\tau_{\rm dyn})$. The proportionality $f_{\rm dyn}=20$ is a free parameter, 
and $\tau_{\rm min}$ is a minimum starburst duration, which is set to $\tau_{\rm min}=100\,\rm Myr$.  
\item The mass entrainment of supernovae driven winds, $\beta$, in {\tt GALFORM} is parametrised by 
the circular velocity of the galaxy, taken to be a proxy for the gravitational potential well. 
In Lacey14, $\beta=(V_{\rm circ}/320{\rm km\,s^{-1}})^{-3.2}$. Detailed calculations 
of SNe feedback suggest the power-law index of this parametrisation 
should be in the range $-1$ to $-2.7$ (\citealt{Murray05}; \citealt{Creasey12}; \citealt{Lagos13}).
 {The power-law index of $-3.2$ in Lacey14 comes from the Monte-Carlo exploration of parameters 
originally done in \citet{Bower06}, which pointed to that value as the best parameter to recover a flat 
faint-end in the $K$-band luminosity function. However, since this value is in tension with the more recent 
studies above, we are currently exploring the effect of including more physical parametrisations 
(\citealt{Mitchell14}; Mitchell et al. in prep.). However, we expect the effect of these new parametrisations 
on the conclusions presented in this paper to be secondary. This is because the selected model ETGs for the study here 
are relatively massive $L_{\rm K}>6\times 10^9\,M_{\odot}$, and more affected by AGN feedback (see Paper I for details).}
\item In {\tt GALFORM} AGN feedback is assumed to act in haloes where the cooling time is longer than the free fall 
time at the cooling radius
(`hot accretion' mode; \citealt{Fanidakis10b}).
In such a halo, the AGN power is computed and if
it is greater than the cooling luminosity, the cooling flow is switched off (see \citealt{Bower06}).
\item For photoionisation feedback, 
it is assumed that no gas is allowed to cool in haloes with a
circular velocity below $V_{\rm crit}$ at redshifts below $z_{\rm
reion}$ \citep{Benson02}. Taking into account simulations by \citet{Okamoto08} and observational
constraints on the reionisation redshift \citep{Komatsu11}, Lacey14 adopt
$V_{\rm crit}=30\,\rm km \,s^{-1}$ and $z_{\rm reion}=10$. 
\item For chemical enrichment, we
 adopt the instantaneous mixing approximation
for metals in the ISM, and change the amount of metals recycled depending on the IMF adopted. 
The stellar evolution models of \citet{Marigo01} and \citet{Portinari98} are adopted 
to calculate the ejected mass and metals from intermediate and massive stars, respectively.
\item Lacey14 adopt the updated dynamical friction timescale 
of \citet{Jiang07} to estimate the timescale for the orbital decay of satellite galaxies towards the centre. This decay is due to energy and angular
momentum losses driven by dynamical friction with the DM halo material. Once the galaxy merger 
takes place, a starburst takes place in the centre, where the SFR is taken to be proportional to the amount of cold gas (HI plus H$_2$)  
in the system, regulated by a SF timescale. This SF timescale is calculated as described above. Note that the \citet{Jiang07} formula is an update 
of the widely used \citet{Lacey93} dynamical friction timescale, using recent $N$-body simulations.
\item Global disk instabilities in galaxies
 occur if the disk becomes sufficiently massive that its self-gravity is dominant over the pressure produced by rotation. 
In this case, the disk becomes 
unstable to small perturbations caused by minor satellites or DM substructures. 
The criterion for instability was described by \citet{Efstathiou82}, \citet{Mo98} and \citet{Cole00}.
SF in the case of instabilities proceeds as in starbursts driven by galaxy mergers.
{It is important to remark here that the way disk instabilities are treated in {\tt GALFORM} are a simplification of what detailed simulations 
 show. For example \citet{Ceverino10}, \citet{Bournaud11} and \citet{Bournaud14} show that when disks are globally unstable, 
large clumps of gas are formed 
that can be long lived due to the large accretion rates that counteract the mass loss due to 
outflows and stripping. These long live clumps can  
migrate to the central parts of the galaxy and build a large bulge in less than $1$~Gyr. This is captured in our simplified model 
through the triggering of a starburst in the central bulge of galaxies. However, contrary to our assumption here, simulations 
show that the disk is not fully destroyed in the process of disk instability.  
We plan to investigate the effect this has on the history of ETGs in a future paper.}
\item The standard treatment of the hot gas in the subhaloes around satellite galaxies is usually referred to as `strangulation' of the hot gas, which
means that once a galaxy becomes a satellite (when it crosses the virial radius of the larger halo), its hot gas is removed instantaneously
and is transferred to the hot gas reservoir of the main halo. This is the model adopted in the default Lacey14 model. 
However, in Paper I we show that 
this treatment is too extreme and leads to ETGs having too low HI and H$_2$ gas fractions compared to observations. 
In Paper I we argue that gradual ram pressure stripping of the hot gas is needed in order to bring gas fractions into 
agreement with observations. The latter is calculated by considering the ram pressure of the hot gas in the main halo acting against the 
hot gas of the satellite galaxy as it moves through the halo,  
and the gravity of the satellite retaining part of the hot halo (see \citealt{McCarthy08}; 
\citealt{Font08}). The Lacey14 model including this scheme of gradual ram pressure stripping of the hot gas is referred to as Lacey14+GRP.
The latter is our standard model for the rest of the paper, which we show in Paper I to provide HI and H$_2$ gas fractions and mass functions 
in very good agreement with the observations of the ATLAS$^{\rm 3D}$ and HIPASS surveys.
{Note that we do not include ram-pressure stripping of the cold gas disk of galaxies. Although this process has been shown 
to be important in cluster environments (e.g. \citealt{Cortese11}; \citealt{Boselli14}), most ETGs in the model and 
in ATLAS$^{\rm 3D}$ reside in environments other than galaxy clusters, and therefore we do not expect this process 
to change the conclusions we present in this paper (see paper I for more details).}
\item To calculate the circular velocity and size of the galaxy disk in {\tt GALFORM} it is assumed that 
there is conservation of angular momentum
and centrifugal equilibrium in the process of gas cooling, while for 
the size of bulges, energy conservation and virial equilibrium are assumed during galaxy mergers or disk instabilities. 
In addition,  
the mass distribution in
the halo and the lengthscales of the disk and the bulge adjust adiabatically in response
to their mutual gravity.
\end{itemize}

\section{Following flips in the angular momentum of the gas, stars and dark matter halo}\label{FlipsModel}

\begin{figure}
\begin{center}
\includegraphics[trim = 1mm 0mm 1mm 0mm,clip,width=0.46\textwidth]{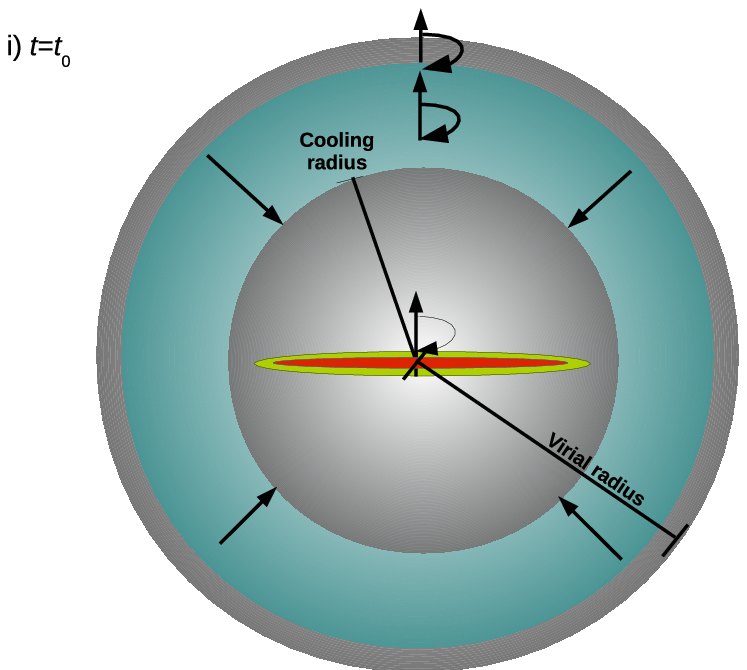}
\includegraphics[trim = 1mm 0mm 1mm 0mm,clip,width=0.46\textwidth]{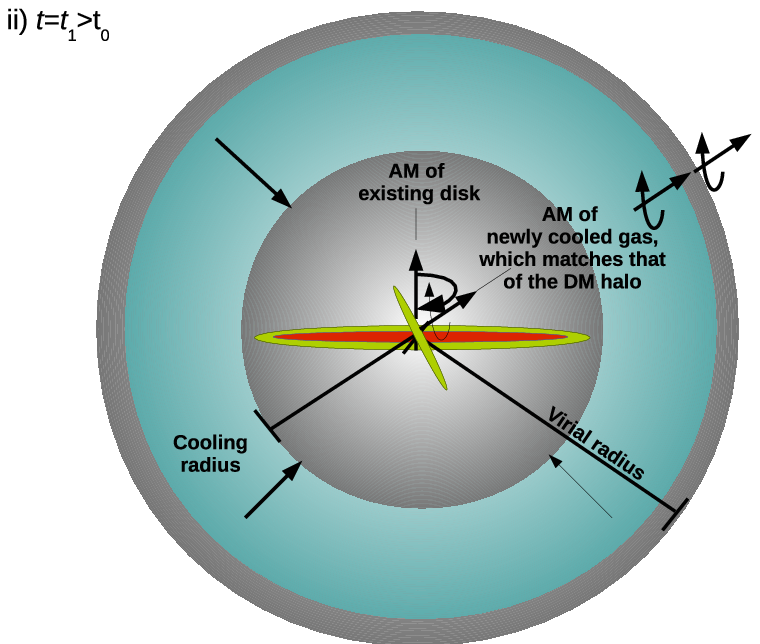}
\includegraphics[trim = 1mm 0mm 1mm 0mm,clip,width=0.46\textwidth]{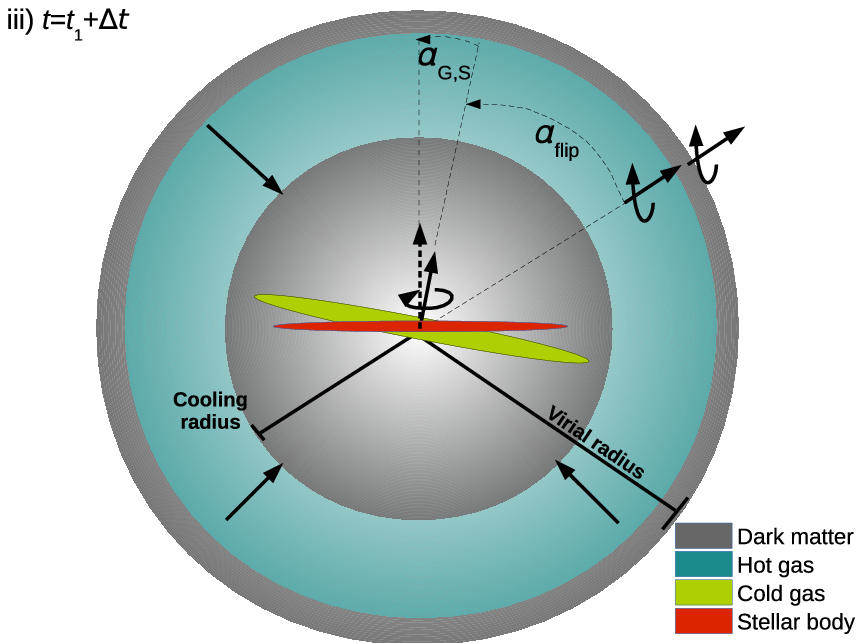}
\caption{Schematic showing the methodology of \citet{Padilla14} applied to our simulated ETGs.
In panel (i) we show a recently formed galaxy, in which all the components have angular momenta that are perfectly aligned. 
In a later timestep, $t_1$ (panel (ii)), the DM halo and hot gas flip their angular momenta direction 
due to accretion. This flip can only be propagated to the galaxy by the accretion of cooled 
gas from the hot halo, which is shown as a condensation of cold gas with a spin aligned with the DM halo. 
The flip the disk suffers is limited by the mass of accreted cooled gas relative to the mass in the disk. 
In panel (iii) we show the case where only a small amount of cooled gas is accreted relative to the gas mass that was already in the disk, 
driving a small flip. 
The outcome of this processes is a misalignment $\alpha_{\rm flip}$ between the gas disk and the dark matter halo and 
an angle $\alpha_{\rm G,S}$ between the gas disk and the stellar body.} 
\label{Misalignmentv2}
\end{center}
\end{figure}

\begin{figure}
\begin{center}
\includegraphics[trim = 1mm 0mm 1mm 0mm,clip,width=0.49\textwidth]{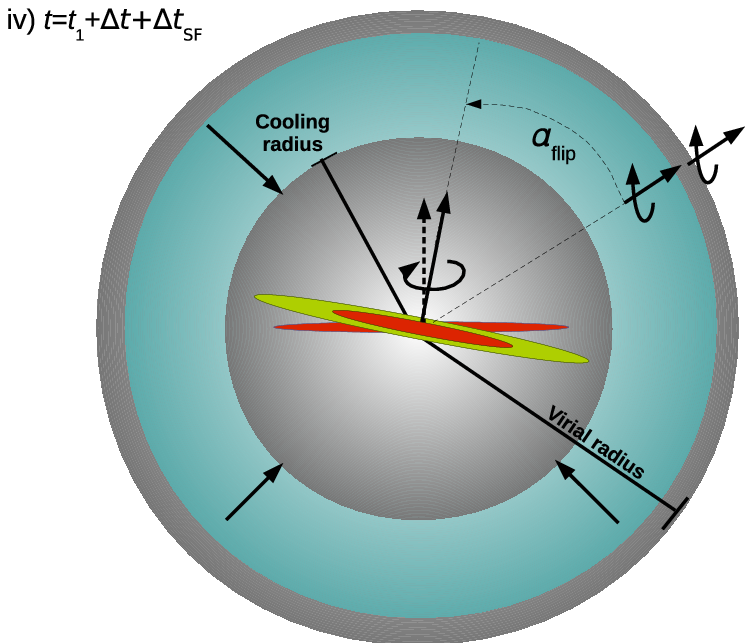}
\includegraphics[trim = 1mm 0mm 1mm 0mm,clip,width=0.49\textwidth]{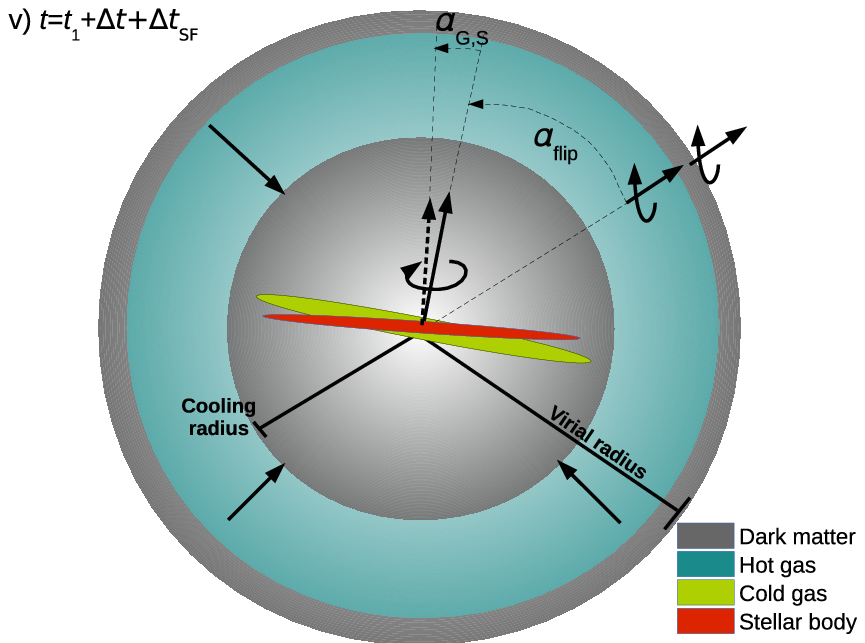}
\caption{Continuation of Fig.~\ref{Misalignmentv2}.
In panel (iv) we show that the cold gas disk has been flipped due to the recent gas accretion. Stars 
form in this flipped gas disk in a timescale $\Delta t_{\rm SF}$. The angular momentum of the stellar disk is shown by the dashed arrow. 
In (v) we show the resulting flip in the stellar disk due to the 
formation of new stars, which had a different angular momentum direction than that of the older 
stars. In this example, the newly formed stars represent a lower mass than the older stars, which 
is not enough to flip the stellar disk back into alignment with the gas disk. Therefore 
there is a residual angle between the angular momenta of the gas disk and the stars, $\alpha_{\rm G,S}$.}
\label{Misalignmentv2cont}
\end{center}
\end{figure}

We use the Monte-Carlo based method developed by \citet{Padilla14} to follow the flips in the 
angular momenta of the cold gas (atomic plus molecular) 
and stars in galaxies, and the DM halo. Padilla et al. based their study on the 
Millennium-II simulation \citep{Boylan-Kolchin09}, which has a resolution that is $125$ 
times better than the Millennium simulation described in $\S$~\ref{modelssec}. 
This allowed Padilla et al. to accurately calculate the angular momentum, $\vec{J}$, in 
all DM haloes and subhaloes with masses $\gtrsim 10^{10}\, h^{-1}\,M_{\odot}$, 
based on at least $1000$ particles\footnote{Note that in Millennium-I a halo of mass 
$10^{10}\, h^{-1}\,M_{\odot}$ is resolved with $10$ particles only, which makes the measurements of angular momentum 
 very noisy}. Thus, the direction 
of $\vec{J}$ can be accurately traced before and after accretion of matter and merger events. Padilla et al. 
built a probability distribution function (PDF) for the angle between $\vec{J}$ 
before and after the accretion, $\Delta\alpha_{\rm DM}$, which depends on 
the relative change in mass. {There are also variations with redshift, due to the expansion of the universe and the relative 
angular momentum brought by the material as time passes. Thus, we build the PDFs for different redshifts and relative mass changes.}
In addition, smooth accretion of matter (which is from the cosmic web and 
it is not resolved in individual haloes) and halo mergers drive different 
magnitudes of angular momentum flips even though they may cause a similar change in mass, with halo mergers 
tending to drive higher $\Delta\alpha_{\rm DM}$ than accretion onto haloes (see Appendix~\ref{PDFsP14} for the PDFs produced by 
Padilla et al.). 
 
The strongest assumption Padilla et al. 
make to apply this methodology to galaxies is that accretion onto galaxies produces flips in the angular momenta, 
 $\Delta\alpha_{\rm bar}$, of the same order as in the haloes (namely, it will depend on the relative change in mass, on the 
nature of the accretion, i.e. whether it is from smooth accretion or mergers, and on time), 
$\Delta\alpha_{\rm bar}(\Delta M_{\rm bar}|M_{\rm bar}, t)\equiv \Delta\alpha_{\rm DM}(\Delta M_{\rm DM}|M_{\rm DM}, t)$. 
In other words, Padilla et al. sample the PDFs created from the Millennium-II simulation to find 
by how much the angular momentum of a galaxy flips due to accretion. This can be done for the different galaxy components.

\citet{Padilla14} implemented the method above in the semi-analytic model {\tt SAG} 
\citep{Lagos08,Jimenez11}, modifying 
 the specific angular momentum of galaxies in the model. This also affects the sizes of 
galaxies and therefore star formation and the strength of feedback. Padilla et al. did not change any model parameters and show 
the effect of this new physical model on galaxy properties. \citet{Ruiz14} presented the 
recalibration of the \citet{Padilla14} model and the fits to the stellar mass function.

Here, 
we apply the \citet{Padilla14} model as post-processing to {\tt GALFORM} 
to calculate the flips in the direction of the angular momentum of ETGs due to their growth 
history, but we do not modify the sizes of galaxies or any other property as a result. For this reason, there is no need 
for changing any parameter in the model. 
By post-processing 
{\tt GALFORM} using the Padilla et al. Monte-Carlo method, we can accumulate the flips of the angular momentum vector of the gas disk 
and, based on the assumption that new stars form perfectly aligned with the gas disk at the time they form, we can calculate the angle 
between the angular momenta of the gas disk and the stellar body at each timestep in the evolution of galaxies.
{Although these changes are expected in such a model, \citet{Padilla14} show that the changes are not major except in the fraction of massive galaxies that are ETGs, in a way that the model including the change in the specific angular momentum
due to flips in the direction of the angular momentum vector produces a higher fraction of ETGs than the previous model. We therefore do not expect a self-consistent calculation to strongly change the conclusions we present in this paper.}

We can take the predicted history of each ETG in our simulation (based on the Lacey14+GRP model) 
and calculate the angle between the gas and stars in galaxies and the DM halo. The history of ETGs is fully 
characterised by minor and major merger events, disk instabilities, accretion of gas from the halo, outflow gas to the halo by 
recent star formation (SNe feedback), mass loss from intermediate and low-mass stars to the ISM, and {heating of the halo gas 
 due to the energy injected there by the AGN (AGN feedback)}. 
All these events will change the gas and stellar contents of ETGs. 

The way we apply 
the Padilla et al. methodology to our galaxies is as follows:

\begin{itemize}
\item In an individual halo, the angular momenta of the hot gas and dark matter are perfectly aligned before 
the first galaxy forms.
\item When the first galaxy forms, the gas cools down preserving the specific angular momentum of the hot gas and its direction, 
leading to a gas disk that is perfectly aligned with the hot halo (see panel (i) in Fig.~\ref{Misalignmentv2}). Since stars form 
from this gas disk, they will also be aligned with the gas and hot halo components. 
\item When a DM halo of mass $M_{\rm DM}$ accretes matter, 
$\Delta M_{\rm DM}$, at a time $t$, it {flips} its angular momentum 
direction by an angle $\Delta\alpha_{\rm DM}(\Delta M_{\rm DM}|M_{\rm DM}, t)$. This change is instantly propagated to the 
angular momentum of the hot gas (see panel (ii) in Fig.~\ref{Misalignmentv2}). 
As a result, there will be some level of misalignment between the gas 
in the disk that cooled from the hot gas before it flipped and the dark matter halo. 
Note that the instantaneous reaction of the hot halo assumed here is for simplicity. 
In reality, there is a torque timescale associated with this process, although this should act on a relatively short timescale of 
the order of the dynamical timescale of the halo. The stars remain fully aligned with the cold gas disk. 
\item When new gas cools down from the flipped hot gas halo, it comes misaligned with the existing gas and stellar disk 
(see panel (ii) in Fig.~\ref{Misalignmentv2}). {The timescale for this gas to cool down and be accreted onto the galaxy disk 
is calculated as described in $\S$~\ref{modelssec} (second bullet point).}
{The cooled gas will change the direction of the angular momentum vector of the cold gas disk towards that of the hot halo 
(see panel (iii) in Fig.~\ref{Misalignmentv2}). 
The resulting cold gas disk will form stars, and will also change the direction of the angular momentum vector of the stellar disk 
towards that of the cold gas disk (as in panel (iv) in Fig.~\ref{Misalignmentv2cont}).
Depending on the cooled mass and the mass of the newly formed stars relative to the existing disk, there will be 
a remaining angle between the angular momenta of the cold gas and stellar disks (panel (v) in Fig.~\ref{Misalignmentv2cont}).
The resulting flip in the angular momentum direction of the cold gas and stellar components are calculated as a mass-weighted angle}. 
From this it is 
implicit that the stars will have an angular momentum direction set mainly by the angular momentum direction of stars at the 
time of the peak of the star formation activity. 
\item When there is a process such as disk instabilities or galaxy mergers that drive the formation of galaxy bulges from a preexisting 
stellar disk, we assume that the newly formed bulge preserves the direction of the angular momentum the stellar disk had. 
Any further star formation episode will affect the direction of the angular momentum of the bulge as a result. 
This means that when we talk about stellar component we mean bulge plus disk, and assume these are always aligned with each other.
 This is supported by theoretical work that shows how the components slew to each other to align their angular momenta  
\citep{Binney86}.
\item During disk instabilities, the cold gas is consumed in a starburst and therefore loses its memory of 
the flips it had accumulated over time, while stars preserve this memory. In a subsequent gas accretion episode 
the gas will come with the angular momentum direction of the current hot halo, and may form stars. We calculate 
the change these newly born stars produce in the angular 
momentum vector of the stellar body as we explained above (see panel (vi) in Fig.~\ref{Misalignmentv2cont}).
\item In the case there is a galaxy merger, we take the PDFs corresponding to the galaxy merger case (see  Appendix~\ref{PDFsP14}), 
and then we apply the same methodology as above to get the new angle between the angular momenta of the cold gas and stellar body. 
For the same fractional change in mass, galaxy mergers tend to produce larger flips in the angular momentum direction than 
results from gas accretion.
\end{itemize}

Since the growth history is recorded for each ETGs at the output times of the simulation, 
we perform the above calculations at every time step, producing a history for the angle between the angular momenta 
of the different components of ETGs. An important assumption in this model is that consequent flips in the angular momenta of 
the galaxy components are not correlated. This means that flips can change the angular momenta in random directions, and are not necessarily 
correlated with the direction at previous times. We test this assumption in $\S$~\ref{limitations}.

Note that in this model we do not take into account any relaxation of the gas disk towards the stellar component due to 
 torques.
In the case of the dark matter component this is implicitly included, as the distributions of flips was obtained 
by \citet{Padilla14} using a $N$-body simulation, which includes the gravitational interactions. If the gas is faster in its 
relaxation than the DM components, then the calculation here would represent upper limits for the angle between the gas and the 
stellar components of ETGs. 
An important consideration is that this is done for individual galaxies in their sub-haloes. This prevents central 
galaxies from changing their angular momentum direction due to substructures in the halo. In other words, 
central galaxies are not affected by changes in the angular momenta of their satellite galaxies (unless they merge).

In observations the measured position angle of the gas is measured in ionised and molecular gas, which are aligned, suggesting 
a common origin \citep{Davis11}. However, in our model we treat the cold gas, which includes atomic and molecular gas, as a single 
component, and therefore we do not distinguish possible misalignments between HI and H$_2$. In observations, HI is not always 
aligned with H$_2$, and indeed a $30$ per cent of ETGs show unsettle HI morphologies \citep{Serra12}, while this percentage in the 
case of H$_2$ is much lower. We simplified the problem by considering the bulk neutral gas and in our model this approximation might 
not be so critical as the neutral gas that is subject to environmental interactions is the lower density gas which does not 
contributes so significantly 
to the total neutral gas mass. 

\section{The history and morphological development of ETGs}

Morphology is a transient property of galaxies that is tightly connected to their growth history. This is because 
it depends on the ability of galaxies to grow galaxy disks after events which lead to spheroid formation, such as disk instabilities 
or galaxy mergers. Throughout the paper we consider ETGs at $z=0$ in the Lacey14+GRP model, which we selected to have  
$L_{\rm K}>6\times10^9\,L_{\rm K}$, $M_{\rm HI+H_2}>10^7,M_{\odot}$ and $M_{\rm bulge}/M_{\rm tot}>0.5$, where 
$L_{\rm K}$ is the $K$-band luminosity, $M_{\rm HI+H_2}$ is the mass of HI plus H$_2$, $M_{\rm bulge}$ is the stellar mass in the bulge and 
$M_{\rm tot}$ is the total stellar mass of the galaxy. These selection criteria are adopted to mimic the selection criteria of  
ATLAS$^{\rm 3D}$ (see Paper I for details).

\begin{figure}
\begin{center}
\includegraphics[trim = 0.9mm 0.3mm 1mm 0.45mm,clip,width=0.49\textwidth]{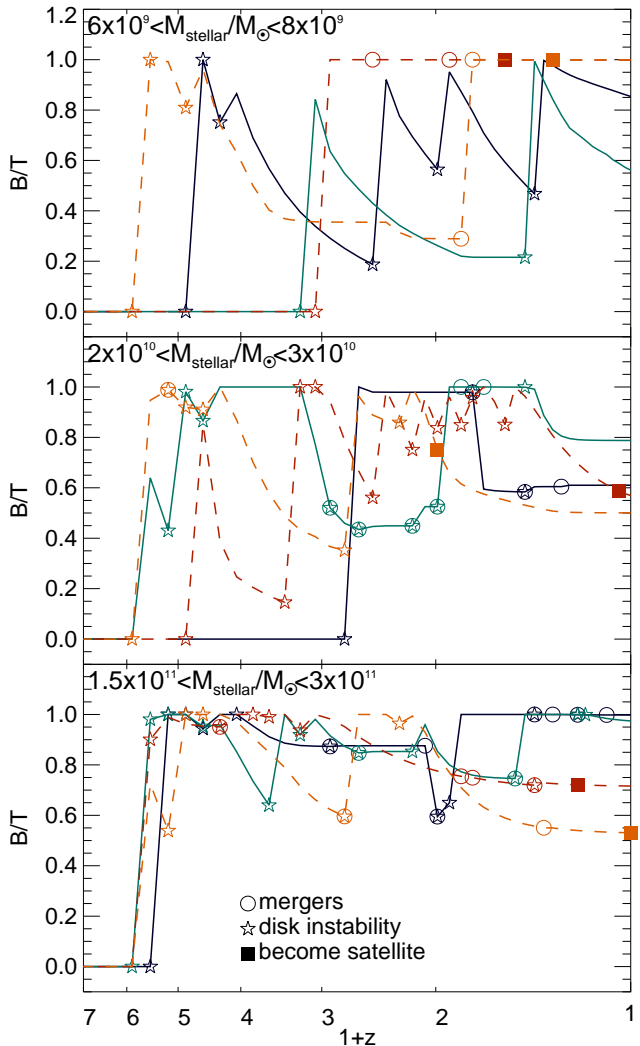}
\caption{Bulge-to-total stellar mass ratio as a function of redshift for $4$ randomly selected galaxies 
with present-day stellar masses in three different ranges, as labelled. 
 Solid lines correspond to randomly selected present-day central galaxies, while dashed lines 
show satellite galaxies. Using symbols we show where each galaxy undergoes a galaxy merger (circles), disk instability (stars) 
and when it becomes satellite (filled square).}
\label{BThistory}
\end{center}
\end{figure}

In order to help visualise the transient nature of morphology, we show in Fig.~\ref{BThistory} the bulge-to-total stellar mass ratio, $B/T$, 
 of randomly selected ETGs in different present-day stellar mass bins, separating central galaxies (solid lines) from 
satellites (dashed lines). The first interesting feature is that massive galaxies tend to have had an 
early-type morphology (i.e. $B/T>0.5$) for a longer 
period than lower mass galaxies. This is seen in the example cases of Fig.~\ref{BThistory}, and also applies to the general ETG 
population. The cause of this difference between massive and lower mass ETGs is that 
low mass galaxies tend to regrow their disk more efficiently than more massive galaxies. 
In the case of the most massive galaxies, this is typically due to AGN feedback preventing the formation of a new disk. This was discussed 
in detail in Paper~I. The more efficient regrowth of galaxy disks in the lower mass ETGs is due to the continuous accretion of 
newly cooled gas. The reason why gas accretion continues even at later times (see, for example, the central galaxy in the top 
panel of Fig.~\ref{BThistory} that has a $B/T=0.56$ at $z=0$) is that these ETGs are hosted by lower mass haloes 
($M_{\rm halo}\lesssim 5\times 10^{11}\,M_{\odot}$) and have no AGN to heat cooling flows. Central galaxies in the middle and bottom panels
of Fig.~\ref{BThistory} are hosted by higher mass haloes, $M_{\rm halo}\gtrsim 10^{12}\,M_{\odot}$. 
We also show in Fig.~\ref{BThistory} the times when galaxies underwent galaxy mergers and disk instabilities,
and in the case of satellite galaxies, we show when they were accreted onto their current halo. 

Note that every time a galaxy in the model displays a rapid growth in its $B/T$ ratio, 
this is due to a galaxy merger or disk instability, as can be seen in the example galaxies in Fig.~\ref{BThistory}. Also note that 
not all galaxy mergers lead to higher $B/T$, but in some cases there is an associated growth of the disk (see for 
example one of the central galaxies in the bottom panel of Fig.~\ref{BThistory} that had a galaxy merger at $z=1.1$). This can happen during minor 
mergers with very small satellite-to-central galaxy mass ratios. In the case of model satellite galaxies, there can be disk regrowth 
 once they become satellites in the case of gradual ram-pressure stripping (which is implemented in the Lacey14+GRP model 
we use here). An example of this are the satellite galaxies in the middle panel of Fig.~\ref{BThistory}, which continue their 
disk regrowth, 
compared to a satellite galaxy in the bottom panel, which stops growing its $B/T$ ratio.
The satellite galaxies in the middle panel of Fig.~\ref{BThistory} 
show regrow of their disks as a consequence of the gradual ram pressure stripping mechanism  
included in the Lacey14+GRP model, as opposed to the strangulation scenario that results in no further gas cooling.
The satellite galaxy of the bottom panel of Fig.~\ref{BThistory}  
that became a satellite at $z\approx 0.3$ does not show any disk regrowth (i.e. its $B/T$ does not
significantly change in the time the galaxy has been satellite). The reason for this is that the halo mass of the latter galaxy 
is higher than those of the satellite galaxies in the middle panel of Fig.~\ref{BThistory}. This translates into ram pressure stripping being 
more efficient in the former case.

Very different stellar mass assembly histories can lead to galaxies having an early-type morphology, as the example galaxies of 
Fig.~\ref{BThistory} show. The specific history 
of each ETG is expected to have a strong effect on the alignment of their stellar, gas and dark matter components. In the model of 
\citet{Padilla14}, every galaxy merger, disk instability and smooth accretion event is considered when following the angular momenta 
flips. We use the individual gas accretion 
history of every ETG to estimate their angular momenta alignments in the next section.

\section{The origin of misaligned gas disks in early-type galaxies}\label{OriginAngle}

In this section we discuss the expected frequency of ETGs 
that have
gas disks misaligned with their stellar component in the Lacey14+GRP model.
We compare extensively with the observations of the ATLAS$^{\rm 3D}$, which are discussed in detail in 
\citet{Davis11}. We find that 
the percentage of misaligned gas disks and stellar components in the ETGs of ATLAS$^{\rm 3D}$ is $\approx 42\pm 6$ per cent\footnote{The error 
in this percentage correspond to a Poisson error.}.

\subsection{Lower limits on the number of misaligned gas disks in ETGs}\label{LowerLims}
\begin{figure}
\begin{center}
\includegraphics[trim = 0mm 0.3mm 1mm 0.45mm,clip,width=0.49\textwidth]{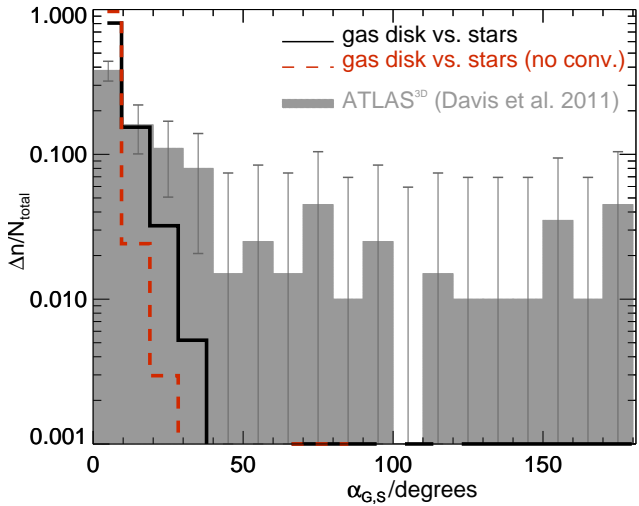}
\caption{Distribution of angles between the angular momenta of the gas disk and the stellar component, $\alpha_{\rm G,S}$ (dashed line), 
 for ETGs with $L_{\rm K}>6\times 10^9,L_{\odot}$ and 
$M_{\rm HI}+M_{\rm H_2}>10^7\,M_{\odot}$ in the 
Lacey14+GRP model, under the test case that the only source of misaligned gas is galaxy mergers. 
Observational measurements from ATLAS$^{\rm 3D}$ are shown by the shaded histogram. Errorbars 
are Poisson errors. The $y$-axis is normalised  
to represent fractions (number of galaxies in the bin divided by the total number in the sample). 
We convolve the predicted distribution of angles with a normalised Gaussian distribution with its
centre at zero degrees and a standard deviation of $15$ degrees (solid line), which corresponds to the expected scatter in the observations 
reported by \citet{Davis11}.}
\label{MisalignmentMergers}
\end{center}
\end{figure}

In order to show how much misalignment is expected 
 from mergers alone we post-process the ETGs selected from the Lacey14+GRP model using the \citet{Padilla14} scheme, but 
we do not apply any misalignment due to gas accretion from the hot halo.
The results of this exercise are shown in Fig.~\ref{MisalignmentMergers}.
If we only allow mergers to drive misalignments, the model fails to reproduce the observed tail of ETGs that have 
angles between the angular momenta of the stars and the cold disk, $\alpha_{\rm G,S}>30$~degrees. We find 
$2$ per cent of ETGs with $\alpha_{\rm G,S}>30$~degrees. This percentage increases to $5$ per cent in the case we adopt 
the \citealt{Lacey93} dynamical friction timescale. This indicates that other sources 
of misaligned gas are important if the model is to agree with the observations.

The fractions calculated above represent lower limits as 
it is assumed that gas accretion onto galaxies from processes other than galaxy mergers come in  
perfectly aligned with the stellar body.
In Paper I, we show that the vast majority of recent galaxy mergers (i.e. those that took place in the last $1$Gyr) 
correspond to minor mergers that drive starbursts only in  
$10$ per cent of cases. The remaining $90$ per cent correspond to the 
accretion of small galaxies onto the larger ETGs. Of all ETGs with cold gas 
masses $M_{\rm HI+H_2}>10^7\,M_{\odot}$, the percentage that have their current cold gas content supplied mainly by 
galaxy mergers ranges from $11$ per cent to $25$ per cent. The main driver of the variation in this percentage is the adopted 
dynamical friction timescale. Adopting the \citet{Jiang07} dynamical friction timescale, as in Lacey14, 
leads to earlier minor mergers and consequently a lower merger rate at $z=0$, leading to the $11$ per cent  
figure referred to above.
 On the other hand, adopting the \citet{Lacey93} dynamical friction timescale, 
as in the models of \citet{Lagos12} and \citet{Gonzalez-Perez13},  
leads to a higher rate of minor mergers at $z=0$, corresponding to the $25$ per cent value referred to above.
However, these mergers do not necessarily cause a notable change in the angular momentum of the stellar and/or 
cold gas bodies. In fact the calculation of \citet{Padilla14} shows that galaxy mergers produce larger angular momentum 
flips than 
other forms of accretion but the angle between the angular momentum of the haloes before and after the merger is 
likely to be small (see Appendix~\ref{PDFsP14}). This angle also depends on the mass ratio between the two galaxies. 
This is the reason why a relatively high fraction of ETGs have their current gas content supplied mainly by galaxy mergers, 
but do not show significant misalignments.

\subsection{The distribution of misaligned gas disks in ETGs due to the history of angular momenta flips}\label{flipsandslews}

We now apply the full scheme introduced by \citet{Padilla14} to {\tt GALFORM} galaxies as a 
more complete calculation than that in $\S$~\ref{LowerLims}. We remind the reader that we take the mass growth history 
of ETGs from the Lacey14+GRP model and post-process it to follow the flips of the angular momenta. 
The full calculation of \citet{Padilla14} includes changes in the direction of the angular momenta of galaxy components
due to both galaxy mergers and the accretion of cooled gas from the hot halo.
 
\begin{figure}
\begin{center}
\includegraphics[trim = 0mm 0.3mm 1mm 0.45mm,clip,width=0.49\textwidth]{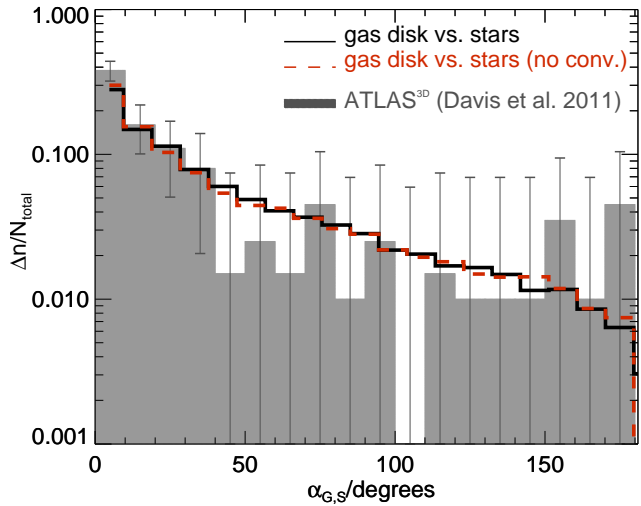}
\caption{Distribution of angles between the angular momenta of the gas disk and the stellar component, $\alpha_{\rm G,S}$ (dashed line), 
and after the convolution with a Gaussian of width $15$ degrees (solid line), for ETGs with $L_{\rm K}>6\times 10^9,L_{\odot}$ and 
$M_{\rm HI}+M_{\rm H_2}>10^7\,M_{\odot}$ in the 
Lacey14+GRP model. Here we fully apply the \citet{Padilla14} Monte-Carlo method.
 The $y$-axis is normalised
to represent fractions (number of galaxies in the bin divided by the total number in the sample). 
Observational measurements are as in Fig.~\ref{MisalignmentMergers}.}
\label{Misalignment}
\end{center}
\end{figure}

Fig.~\ref{Misalignment} shows the predicted distribution function of the angle between the cold gas (HI+H$_2$) and the stellar 
component of ETGs, $\alpha_{\rm G,S}$. 
We also show these distributions after convolving 
with a Gaussian distribution with its
centre at zero degrees and a standard deviation of $15$ degrees, which 
 corresponds to the expected scatter in the observations of \citep{Davis11}.
 Observations of the ATLAS$^{\rm 3D}$ 
reported by \citet{Davis11} are shown by the solid histogram with the error bars corresponding to Poission errors.
The application of the angular momentum flips model described in $\S$~\ref{FlipsModel} leads to $46$ per cent of ETGs 
having $\alpha_{\rm G,S}>30$~degrees and a distribution 
of $\alpha_{\rm G,S}$ that agrees with the observed distribution of ETGs within the error bars (in $\S$~\ref{limitations} we quantify 
the agreement between the model predictions and observations using the Kolmogorov-Smirnov test). This is interpreted 
as misalignment arising not only from recent mergers, as was assumed in $\S$~\ref{LowerLims}, 
but also from the accretion history of ETGs, which can 
 slew the angular momentum of the stellar component over time with respect to the DM halo and also the gas. 

\begin{figure}
\begin{center}
\includegraphics[trim = 2.5mm 0.3mm 1mm 0.45mm,clip,width=0.45\textwidth]{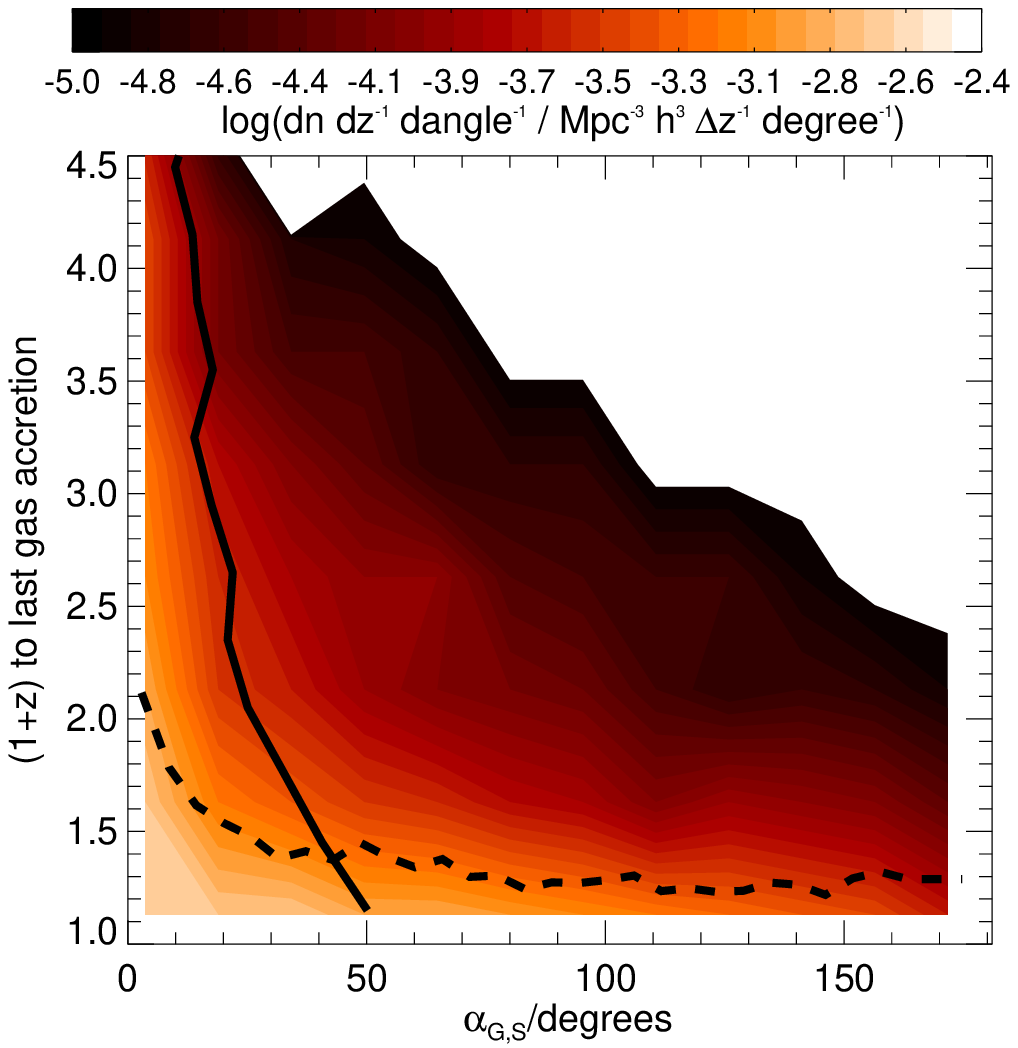}
\includegraphics[trim = 2.5mm 0.3mm 1mm 0.45mm,clip,width=0.45\textwidth]{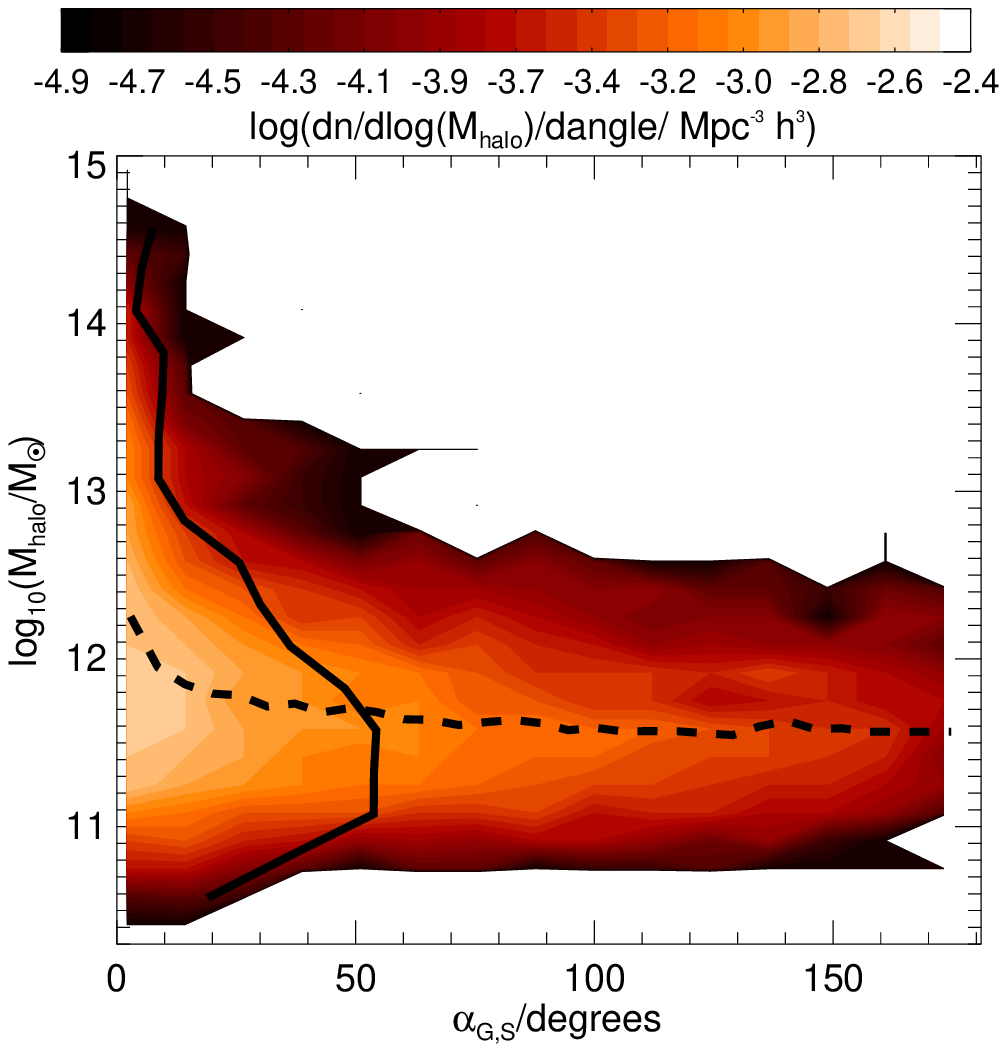}
\caption{{\it Top panel:} The redshift to the last time each ETG had a 
gas accretion episode (either due to gas cooling from the hot halo or galaxy mergers), $z_{\rm LG}$, as a function 
 $\alpha_{\rm G,S}$ in the 
Lacey14+GRP model. Here we include the same model galaxies as Fig.~\ref{Misalignment}. The coloured contours show number density per unit 
redshift and angle, 
with the scale shown by the key ({here d$z$ and d$angle$ are the bins in redshift and angle, respectively}). 
The solid line shows the median $\alpha_{\rm G,S}$ for bins in 
$z_{\rm LG}$, while the dashed line shows the median $z_{\rm LG}$ for bins in $\alpha_{\rm G,S}$. 
{\it Bottom panel:} 
As in the top panel but here we show the sub-halo mass where the ETG resides. {Here 
dlog$(M_{\rm halo})$ and d$angle$ are the bins in halo mass and angle, respectively}.}
\label{LBTdoubling}
\end{center}
\end{figure}

We find that the angle $\alpha_{\rm G,S}$ of each galaxy is strongly affected by recent accretion onto the galaxy compared to the 
time when the galaxy assembled its stellar mass\footnote{We define this as the time the galaxy had assembled 
$80$ per cent of its $z=0$ stellar mass.}. We find that 
 if galaxies assembled $80$ per cent of their stellar mass   
before the last episode of cold gas accretion (either from gas cooling from the hot halo or galaxy mergers), they are 
more likely to show misalignments between the angular momenta of the cold gas and the stellar component. 
Indeed, we find that ETGs that have $\alpha_{\rm G,S}>30$~degrees at $z=0$ assembled $80$ per cent of their stellar mass 
by $z\approx 0.6$, while their last important gas accretion episode took place later at $z\approx 0.3$, on average.
On the other hand, in the sample of ETGs with $\alpha_{\rm G,S}<30$~degrees at $z=0$, we find that $80$ per cent of their stellar mass was 
assembled by $z\approx 0.7$, while their last important gas accretion episode took place at around the same time, on average.
This happens because the new gas accretion episodes do not contribute strongly to the stellar mass, and so they hardly change 
the direction of the angular momentum of stars, while the cold gas can be greatly affected if the change in cold gas mass is important. 

This result is shown in the top panel of Fig.~\ref{LBTdoubling}, where we show 
the redshift of the last cold gas accretion episode of ETGs (either due to gas cooling from the hot halo or a galaxy merger), 
$z_{\rm LG}$,  
as a function of $\alpha_{\rm G,S}$. This cold gas accretion episode can have any mass, but it is on average quite 
significant; the median 
cold mass accreted in this last accretion episode is $3\times 10^9\,M_{\odot}$ for ETGs with $M_{\rm stellar}=10^{10}\,M_{\odot}$ and 
$7\times 10^8\,M_{\odot}$ for ETGs with $M_{\rm stellar}=10^{11}\,M_{\odot}$.   
We also show the median $\alpha_{\rm G,S}$ for bins in 
$z_{\rm LG}$ and vice-versa to show the dominant trends. 
Recent accretion can lead either to aligned or misaligned cold gas/stellar components, which can be  seen 
from the higher density of galaxies at low $z_{\rm LG}$ regardless of the value of $\alpha_{\rm G,S}$. All ETGs with 
earlier last cold gas accretion episodes have more aligned stellar and gas components. This is clear from the median 
$\alpha_{\rm G,S}\approx 10$~degrees of galaxies that have $z_{\rm LG}>3$. Similarly, the median 
value of $z_{\rm LG}\approx 1$ for galaxies with $\alpha_{\rm G,S}\sim 0$ is higher than for galaxies with higher values of $\alpha_{\rm G,S}$. 
Note that the median $\alpha_{\rm G,S}$ increases 
for decreasing $z_{\rm LG}$.

Another important property that is strongly correlated with the formation time and accretion history of galaxies, 
and therefore with $\alpha_{\rm G,S}$, is the mass of the sub-halo that hosts an ETG.
 The bottom panel of Fig.~\ref{LBTdoubling} shows that misalignment happens in our model 
only in intermediate ranges of sub-halo masses, $7\times 10^{10}\,M_{\odot}<M_{\rm subhalo}<10^{13}\,M_{\odot}$. The main halo
 is the largest gravitationally bound structure, which are identify in the $N$-body simulation using 
the friends-of-friends algorithm \citet{Davis85}. Halos can contain substructures that are self-gravitating, which are call 
sub-haloes and are identify using sub-finder algorithms \citep{Knebe13} (see \citealt{Jiang14} for a recent description on how 
haloes and sub-haloes are identified and followed in the $N$-body simulations used by {\tt GALFORM}). The most massive sub-halo within haloes 
generally hosts the central galaxy, while lower mass sub-haloes contain satellite galaxies.
Not all satellites retain information about their sub-halo given that 
the dark matter halo can get heavily stripped so that is not considered a substructure within the main halo anymore.
 In these cases the sub-halo mass corresponds 
to the mass of the sub-halo the last time it was identified in the simulation. In the case of ETGs living in massive sub-haloes, 
$M_{\rm subhalo}>10^{13}\,M_{\odot}$,   
most of the gas accretion took place at higher redshifts, where most of the star formation happened. This leads to alignment between 
the angular momenta of the stars and the gas; for $M_{\rm subhalo}>10^{13}\,M_{\odot}$, the median 
$\alpha_{\rm G,S}\approx 8$~degrees, while for a halo mass of $M_{\rm subhalo}\approx 3\times 10^{11}\,M_{\odot}$ 
the median $\alpha_{\rm G,S}\approx 50$~degrees.
Similarly, the median halo mass of low $\alpha_{\rm G,S}$ is higher than for higher values of $\alpha_{\rm G,S}$.
Large misalignments, for example $\alpha_{\rm G,S}>90$~degrees, are exclusive of ETGs 
hosted by haloes $7\times 10^{10}\,M_{\odot}<M_{\rm subhalo}<3\times 10^{12}\,M_{\odot}$.

In the case of the lowest mass sub-haloes, 
galaxies have undergone a quieter history, with fewer interactions relative to 
ETGs hosted by larger mass haloes. This leads the ETG population hosted by low mass sub-haloes to having 
 fewer ETGs with $\alpha_{\rm G,S}>30$~degrees than is the case of ETGs hosted by higher mass sub-haloes.

\subsubsection{Variance from field to field}\label{Variance}
\begin{figure}
\begin{center}
\includegraphics[trim = 0mm 0.3mm 1mm 0.45mm,clip,width=0.49\textwidth]{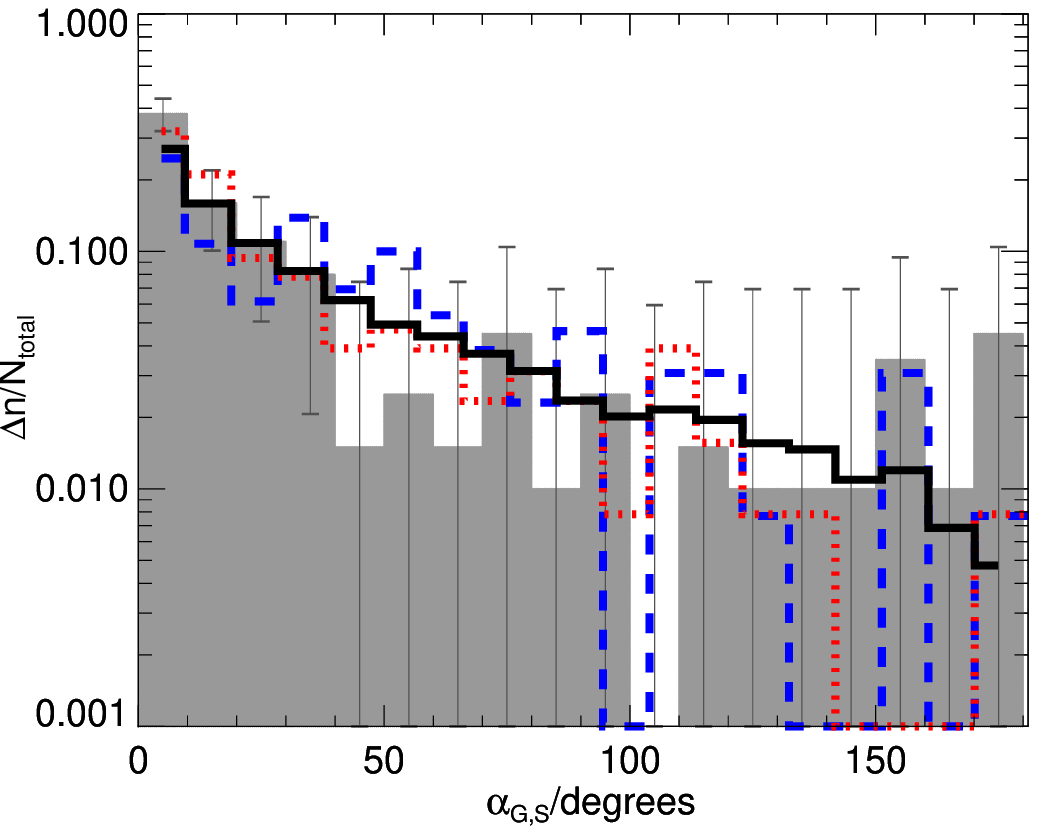}
\includegraphics[trim = 0mm 0.3mm 1mm 0.45mm,clip,width=0.49\textwidth]{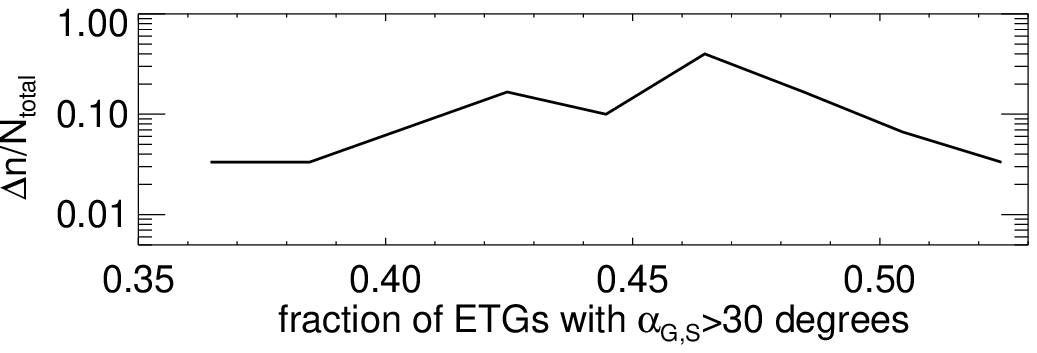}
\caption{{\it Top panel:} Distribution of $\alpha_{\rm G,S}$ 
for ETGs with $L_{\rm K}>6\times 10^9,L_{\odot}$ and
$M_{\rm HI}+M_{\rm H_2}>10^7\,M_{\odot}$ in the
Lacey14+GRP model, and for the realisations of the ATLAS$^{\rm 3D}$ volume ($3\times 10^5\,\rm Mpc^3$) 
that give the highest (dashed line), median (solid line) 
and lowest (dot-dashed line) fraction of ETGs with $\alpha_{\rm G,S}>30$~degrees.
The observed ATLAS$^{\rm 3D}$ distribution is shown by the solid histogram.
 {\it Bottom panel:} 
Distribution of the fraction of ETGs that have $\alpha_{\rm G,S}>30$~degrees in the $30$ realisations done of the 
ATLAS$^{\rm 3D}$ volume in the Lacey14+GRP model.
 The median of the distribution is $46$ per cent.}
\label{MisalignmentVar}
\end{center}
\end{figure}

An important question that arises from a sample such as ATLAS$^{\rm 3D}$, which is 
complete up to a distance of $42$~Mpc, is what is the effect of the large scale structure on the statistical analysis we have 
performed in $\S$~\ref{flipsandslews} (i.e. cosmic variance)? In other words, if we were to observe different volumes of 
$3\times 10^5\,\rm Mpc^3$ (which corresponds to a cube of side $\approx 68$~Mpc), how much variance 
would we observe in the distribution of $\alpha_{\rm G,S}$?
In this section we investigate the variance of the distribution of $\alpha_{\rm G,S}$.

We make a random selection of $30$ sub-volumes within the MS-W7 simulation, with each having a volume of $3\times 10^5\,\rm Mpc^3$.
The resulting distribution of this experiment and the fraction of ETGs with $\alpha_{\rm G,S}>30$~degrees that result 
are shown in Fig.~\ref{MisalignmentVar}. We find that 
 the percentage of ETGs that have $\alpha_{\rm G,S}>30$~degrees can vary from $35-53$ per cent, with a median 
of $46$ per cent. This translates into a variance of $\approx 25$ per cent in the volume probed by ATLAS$^{\rm 3D}$. This number 
agrees very well with the variance as a function of probed volume reported by \citet{Driver10}. 

We also find that the number of ETGs selected to have $L_{\rm K}>6\times 10^9\,L_{\odot}$ and 
a mass of HI plus H$_2$ $>10^7\,M_{\odot}$ is on average 
$~300$ in each sub-volume of $3\times 10^5\,\rm Mpc^3$. This number is very close to the number $260$ of ETGs in the ATLAS$^{\rm 3D}$. 
These variations are quite large and we find that the observed sample of ETGs would need to be extended up to 
$100$~Mpc to reduce the variance in the fraction of ETGs with $\alpha_{\rm G,S}>30$~degrees to a few percent.

\subsubsection{Dependence on galaxy properties}\label{DependencyProps}

In this section we focus on the galaxy properties we find are correlated with 
$\alpha_{\rm G,S}$ and analyse the physical drivers of such correlations. These properties are 
 $K$-band Luminosity and stellar mass, cold gas fraction, satellite/central dichotomy and 
the SFR.
We find no strong dependence of the distribution of 
$\alpha_{\rm G,S}$ on other galaxy properties, such as bulge-to-total stellar mass ratio or  
total neutral gas mass (HI plus H$_2$). 

\begin{figure}
\begin{center}
\includegraphics[trim = 2.5mm 0.3mm 1mm 0.45mm,clip,width=0.49\textwidth]{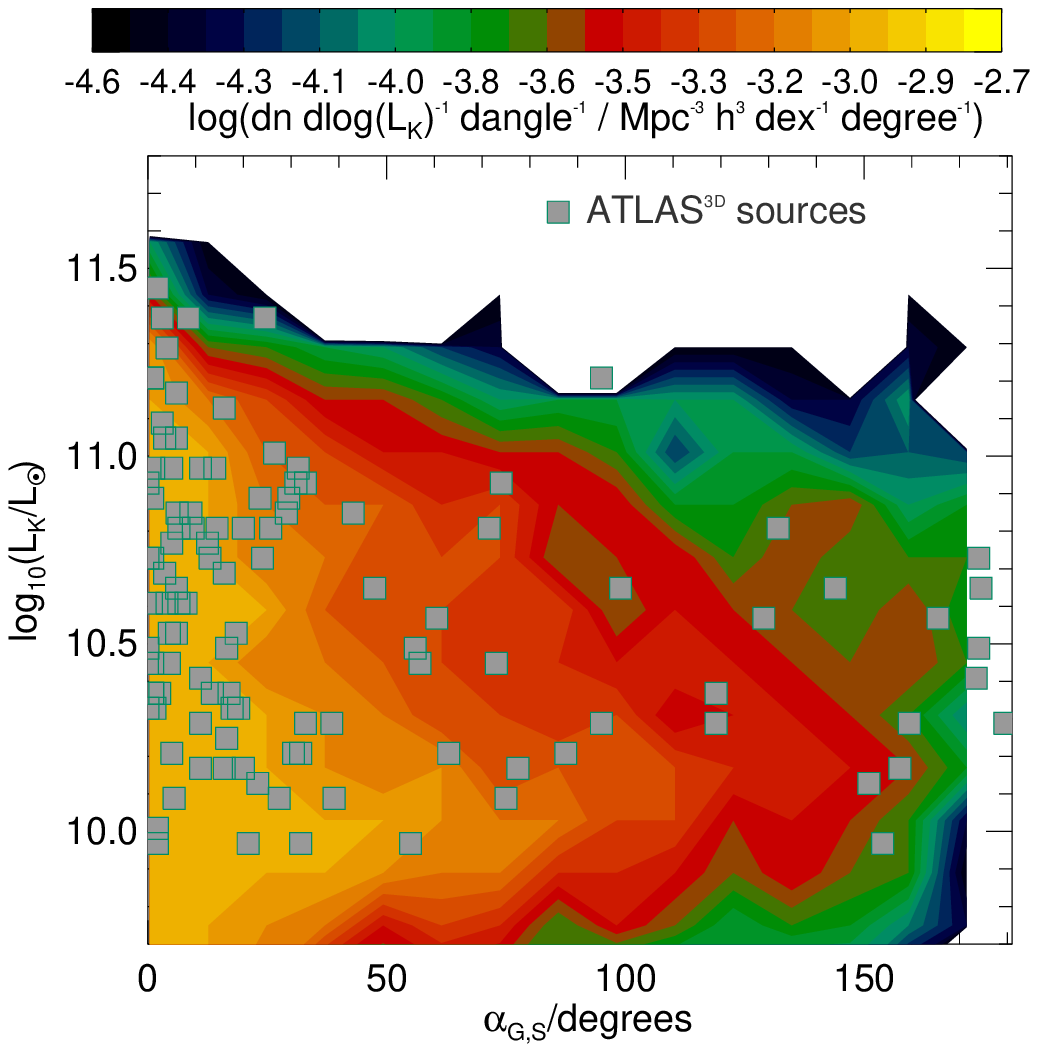}
\includegraphics[trim = 0mm 0.3mm 1mm 0.45mm,clip,width=0.49\textwidth]{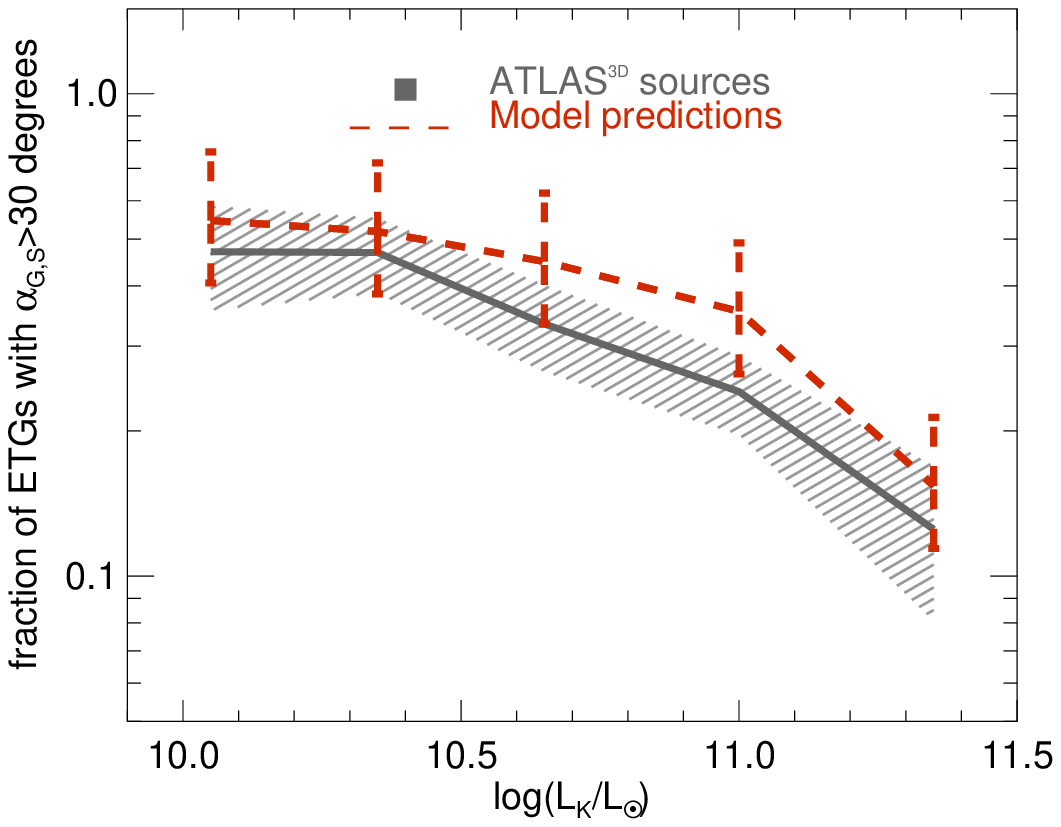}
\caption{{\it Top panel:} $K$-band luminosity as a function of the angle 
between the neutral gas and stellar component, $\alpha_{\rm G,S}$, for  
simulated ETGs from the Lacey14+GRP model and observations from the ATLAS$^{\rm 3D}$ survey \citep{Davis11}. The contours show number density 
per unit $K$-band luminosity and angle, as indicated by the colour bar. 
{ Here
dlog$(L_{\rm K})$ and d$angle$ are the bins in $K$-band luminosity and angle, respectively.}
{\it Bottom panel:} The fraction of ETGs that have $\alpha_{\rm G,S}>30$~degrees
as a function of the $K$-band luminosity. Observational measurements from the ATLAS$^{\rm 3D}$ are shown by the solid line
and hatched region, where the latter corresponds to the Poisson error bars. In the model predictions, the error bars
correspond to the variations in the fraction due to variance from field-to-field (which we discuss in $\S$~\ref{Variance}).}
\label{MisalignmentGas1}
\end{center}
\end{figure}

{\it $K$-band Luminosity and Stellar mass.} 
We find that $K$-band luminosity (often used as a proxy for stellar mass) is anti-correlated with  
the angle between the angular momenta of the cold gas and the stellar component, $\alpha_{\rm G,S}$. The top-panel of 
Fig.~\ref{MisalignmentGas1} shows the 
$K$-band luminosity as a function of $\alpha_{\rm G,S}$. Overlaid are observations of individual galaxies 
from the ATLAS$^{\rm 3D}$ survey \citep{Davis11}. 

The distribution of $\alpha_{\rm G,S}$ seen in the ATLAS$^{\rm 3D}$ observations is very similar to the predictions of the model:
there is a higher number density of galaxies at $\alpha_{\rm G,S}<30$~degrees, with the density of galaxies decreasing towards 
 $\alpha_{\rm G,S}>30$~degrees, but with galaxies covering the full range of $\alpha_{\rm G,S}$ 
at $L_{\rm K}<2\times 10^{11}\,L_{\odot}$. At higher $K$-band luminosities, the model predicts that ETGs show aligned 
gas and stellar components, with the largest angles being $\alpha_{\rm G,S}\approx 20$~degrees, 
in agreement with the observations. These galaxies reside in the highest mass haloes, with a median 
sub-halo mass of $M_{\rm halo}\approx 9\times 10^{13}\,M_{\odot}$ (shown in Fig.~\ref{LBTdoubling}), 
which explains why the model predicts a correlation between $K$-band luminosity and 
$\alpha_{\rm G,S}$. 
For instance, ETGs with $6\times 10^9L_{\odot}<L_{\rm K}<10^{10}L_{\odot}$, have on average its last cold gas accretion episode 
at $z\approx 0.3$, while $80$ per cent of the stellar mass of these ETGs was in place at $z\approx 0.4$. 
In contrast, ETGs with $L_{\rm K}>3\times 10^{11}L_{\odot}$ had the last cold gas accretion episode on average at $z\approx 2.2$, 
while $80$ per cent of the stellar mass of these ETGs was in place only by $z\approx 0.2$\footnote{Note that the timescale we discuss here 
for the stellar mass build-up is an assembly age rather than a stellar age. This is because the stellar ages of these galaxies 
are much higher than the assembly age (see for example \citealt{Kauffmann96} and \citealt{Baugh96}).}. The latter is due to AGN feedback acting 
on the most massive galaxies from early-on. We have repeated this experiment with stellar mass rather than $K$-band luminosity and find 
the same trends. 

In order to quantify the agreement with the observations, we 
calculate the fraction of ETGs in the model and the observations that have 
$\alpha_{\rm G,S}<30$~degrees in bins of $K$-band luminosity. The results of this experiment are shown in the bottom panel of  
Fig.~\ref{MisalignmentGas1}. For the observations, we calculate Poisson error bars, which are shown by the hatched 
region, while for the simulations we calculate an errorbar associated with the variance expected from field-to-field, 
which we discuss in $\S$~\ref{Variance}. There is good agreement between the observations and the model.
Previous calculations by \citet{Dubois14} point to stellar mass being strongly correlated with the spin of 
galaxies and hence is a good indicator of whether the angular momentum of the galaxy is aligned with the halo. However, Dubois et al. do 
not analyse the alignments between the different galaxy components.  

\begin{figure}
\begin{center}
\includegraphics[trim = 0mm 0.3mm 1mm 0.45mm,clip,width=0.49\textwidth]{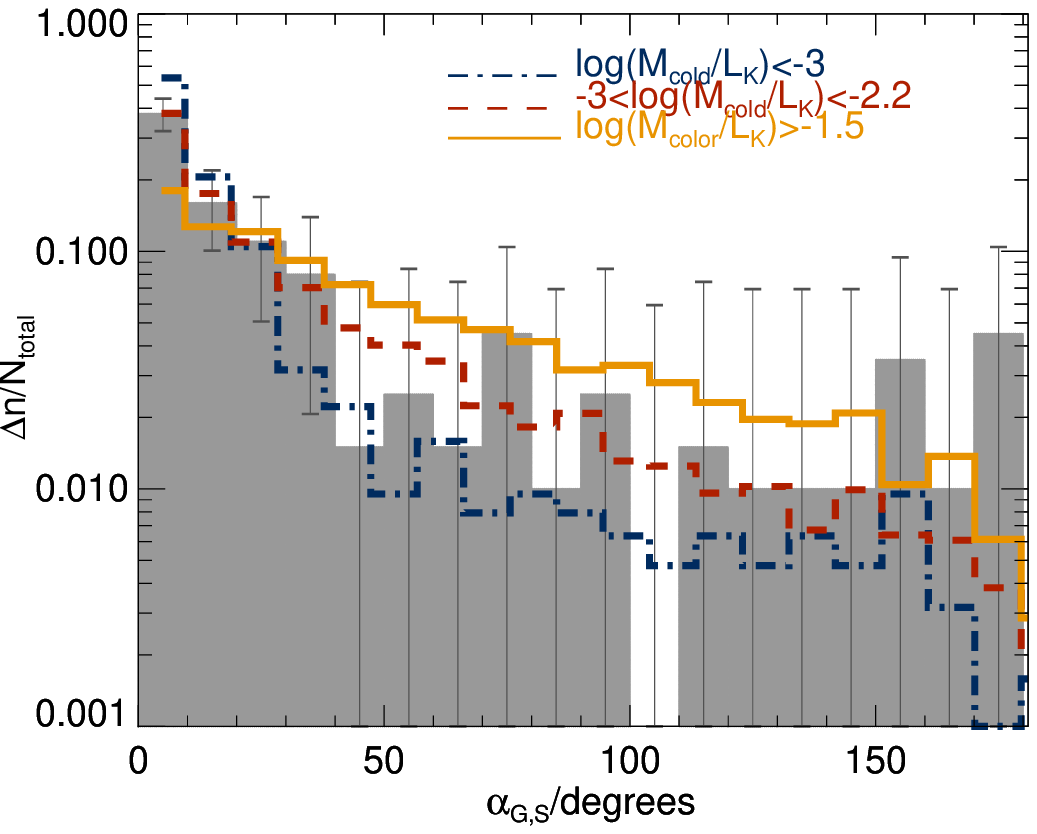}
\includegraphics[trim = 0mm 0.3mm 1mm 0.45mm,clip,width=0.49\textwidth]{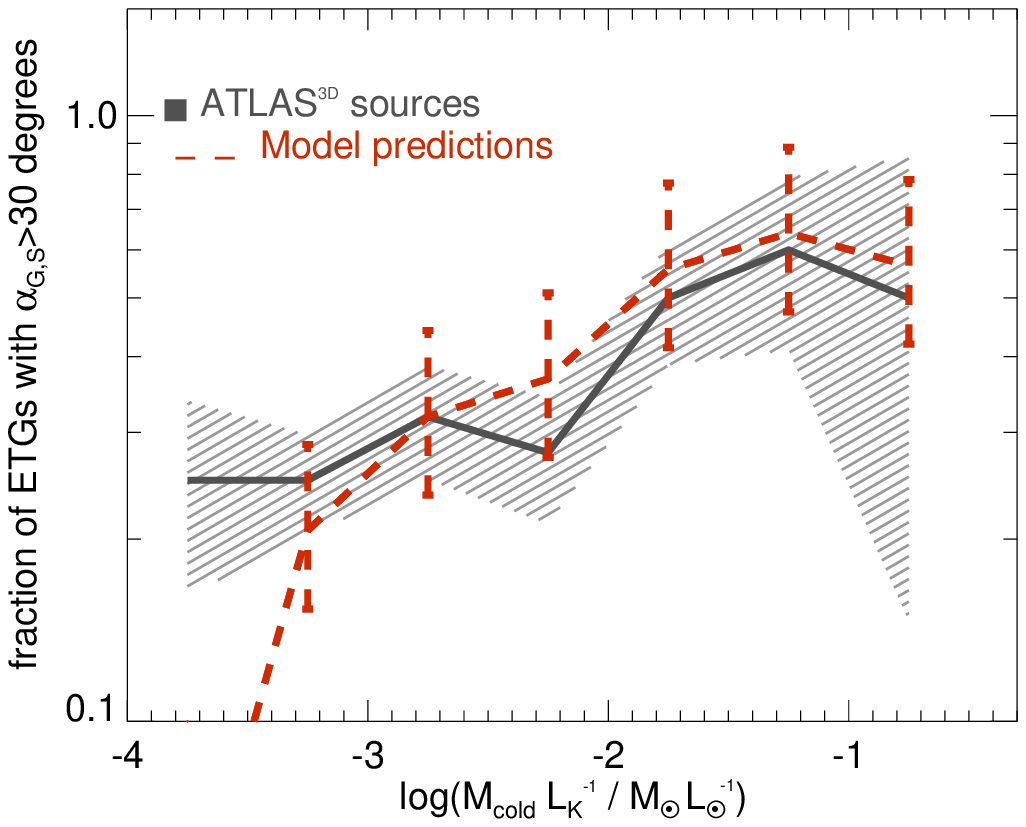}
\caption{{\it Top panel:} Distribution of $\alpha_{\rm G,S}$   
in ETGs with $L_{\rm K}>6\times 10^9\,L_{\odot}$ in the 
Lacey14+GRP model and in different bins of cold gas fraction, $M_{\rm HI+H_2}/L_{\rm K}[M_{\odot}/L_{\odot}]$.
Observational measurements from the ATLAS$^{\rm 3D}$ are shown by the shaded histogram. The $y$-axis is normalised
to represent fractions. {\it Bottom panel:} As in the bottom panel of Fig.~\ref{MisalignmentGas1}, but here 
we show the fraction of ETGs that have $\alpha_{\rm G,S}>30$~degrees 
as a function of the cold gas fraction.}
\label{MisalignmentGasGasFRac}
\end{center}
\end{figure}

\begin{figure}
\begin{center}
\includegraphics[trim = 0mm 0.3mm 1mm 0.45mm,clip,width=0.49\textwidth]{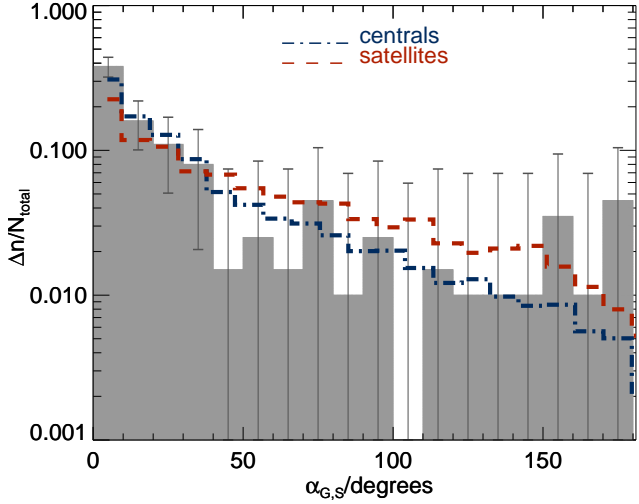}
\caption{As in the top panel of Fig.~\ref{MisalignmentGasGasFRac} but for central and satellite galaxies.}
\label{MisalignmentGasSatCen}
\end{center}
\end{figure}

\begin{figure}
\begin{center}
\includegraphics[trim = 0mm 0.3mm 1mm 0.45mm,clip,width=0.49\textwidth]{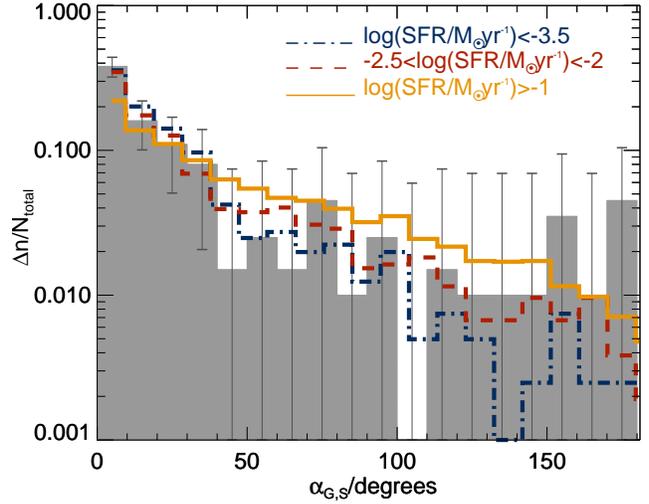}
\includegraphics[trim = 0mm 0.3mm 1mm 0.45mm,clip,width=0.49\textwidth]{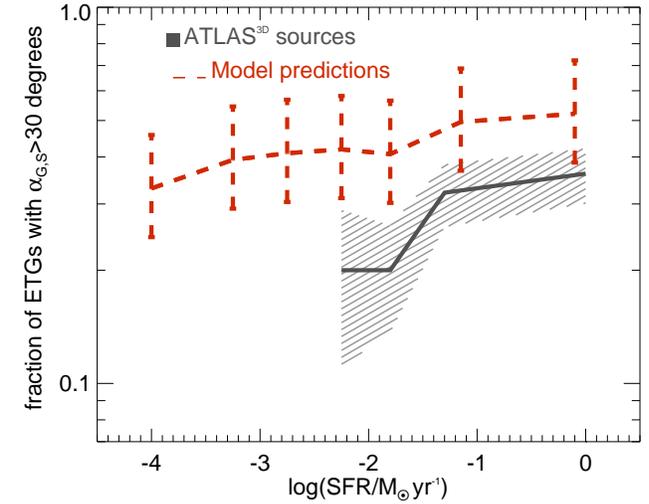}
\caption{As in Fig.~\ref{MisalignmentGasGasFRac} but for bins of SFR. In  ATLAS$^{\rm 3D}$, 
there are SFRs available {from WISE, and combined with GALEX for galaxies that have detection in the far-UV, for only $56$ }
sources \citep{Davis14}. 
This is why the statistics worsen when plotted as a function of the SFR compared to the trends shown in 
the bottom panel of Fig.~\ref{MisalignmentGasGasFRac}.}
\label{MisalignmentGasSFR}
\end{center}
\end{figure}

{\it Gas Fraction.} We find that the cold gas fraction, expressed as 
the ratio between the HI+H$_2$ mass and the $K$-band luminosity, 
 is correlated with the level of misalignment between the angular momenta of the cold gas disk and the stellar component such 
that galaxies with low cold gas fractions tend to be more aligned than galaxies that are gas-rich. This is shown in 
the top panel of Fig.~\ref{MisalignmentGasGasFRac}. 
Galaxies with high cold gas fractions typically have had 
recent cold gas accretion, which have changed the cold gas content considerably, while their 
stellar mass has not changed significantly over the same timescale. The effect of this is that the angular momentum of the 
stellar body is only mildly modified by the newly formed stars, while the angular momentum of the cold gas changes considerably.
The gas fraction here is correlated with the recent gas accretion history. {We find that this relation holds even when we look at 
galaxies in a narrow range of stellar masses, pointing to this correlation being independent of stellar mass.}

\citet{Serra14} show that 
there is a trend of galaxies with low HI fractions to show alignments between the stellar and HI components, while 
misalignments are more likely in ETGs with higher HI gas fractions. This trend agrees qualitatively with our predictions.
We quantify this trend by calculating the fraction 
of ETGs that have $\alpha_{\rm G,S}>30$~degrees as a function of cold gas fraction, (HI+H$_2$)/$L_{\rm K}$, 
for ETGs in both simulations and observations. In the case of non-detections of HI and/or CO for ATLAS$^{\rm 3D}$ galaxies, 
we take the value of the upper 
limit on the HI and/or H$_2$ mass. The results are shown in the bottom panel of Fig.~\ref{MisalignmentGasGasFRac}, where 
the error bars are as in the bottom panel of Fig.~\ref{MisalignmentGas1}.
 We find that the correlation predicted by the model between $\alpha_{\rm G,S}$ and cold gas fraction is also observed 
in the ATLAS$^{\rm 3D}$, which is reassuring. This is evidence that our model contains some of the mechanisms 
that drive the misalignments observed 
in ATLAS$^{\rm 3D}$. 

{\it Satellites vs. Centrals.} Fig.~\ref{MisalignmentGasSatCen} shows the histogram of the angle 
between the angular momenta of the cold gas and stellar components in ETGs, separating central and satellite galaxies. 
There is a trend for satellite galaxies to show misalignments more often than centrals. 
Satellite galaxies in the model are quenched by ram pressure stripping of the hot gas and SNe driven winds. The former 
start operating when the galaxy becomes a satellite. Overall, this results in satellite galaxies having $80$ per cent of 
their stellar mass in place earlier than central galaxies, which exhibit more recent star formation. For satellites, accretion of 
cold gas onto the galaxy can continue for as long as the satellite retains part of its hot gas halo. However, a lot of this cold gas 
is too low density to transform into H$_2$, resulting in low SFRs that do not alter the stellar mass of the galaxy. 
In the case of central galaxies, 
both star formation and accretion of cold gas continue for longer which leads to higher levels of alignment between the two components. 
Unfortunately, we cannot test this prediction with the current available data, but with larger datasets than  
ATLAS$^{\rm 3D}$ we would be able to cross-match with group catalogues and test this idea. 

Note that the trend of satellite/centrals is not driven by stellar mass. We can see this by fixing the stellar mass and then looking at 
satellite and central galaxies. As an example, for the stellar mass bin $10^{10}M_{\odot}<M_{\rm stellar}<3\times 10^{10}M_{\odot}$, 
the fraction of satellite ETGs with $\alpha_{\rm G,S}>30$~degrees is $62$ per cent, while for centrals this fraction is $45$ per cent. 
However, there is a strong connection between the frequency of misalignments in centrals and satellites and the 
range of sub-halo masses hosting these two galaxy populations. The median sub-halo mass of satellite ETGs in the model 
is $M_{\rm subhalo}\approx 3\times 10^{11}\,M_{\odot}$, where the peak of the frequency of misalignments takes place (see bottom panel 
of Fig.~\ref{LBTdoubling}). On the other hand, the median sub-halo mass of central ETGs in the model is 
$M_{\rm subhalo}\approx 10^{12}\,M_{\odot}$.

{\it Star formation rate.} Fig.~\ref{MisalignmentGasSFR} shows the distribution of $\alpha_{\rm G,S}$ 
in three different bins of SFR. We find that in the model, ETGs with low SFRs are associated 
with a preference for lower values of $\alpha_{\rm G,S}$ compared to those with higher SFRs. This is because  
higher SFRs in ETGs are associated with important recent cold gas accretion. 
Note that a similar trend is found between $\alpha_{\rm G,S}$ and the specific SFR (the 
ratio between the SFR and the stellar mass; SSFR); i.e. galaxies with lower SSFR show lower values of $\alpha_{\rm G,S}$.
{At a fixed stellar mass we find that the correlation between $\alpha_{\rm G,S}$ and SFR becomes stronger. This is because 
stellar mass is positively correlated with SFR but anti-correlated with $\alpha_{\rm G,S}$. Thus, when plotting 
all stellar masses the positive correlation between SFR and $\alpha_{\rm G,S}$ becomes weaker. Note however, that the driver of the 
correlation of $\alpha_{\rm G,S}$ with SFR and cold gas fraction is the same and therefore they are not independent.}

We test the existence of a correlation between $\alpha_{\rm G,S}$ and 
SFR using ATLAS$^{\rm 3D}$ and the measured SFRs presented by \citet{Davis14}. 
{Davis et al. measured SFRs in ETGs using a combination of ultraviolet information from GALEX and 
 infrared photometry from WISE, when far-UV photometry was available, or only WISE data when
 not far-UV was available. This was possible for a subsample of $56$ ETGs}. This worsens the statistics, but nevertheless we perform 
the calculation of the fraction of ETGs with $\alpha_{\rm G,S}>30$~degrees as a function of SFR
to reveal possible trends. This is shown in the bottom panel of Fig.~\ref{MisalignmentGasSFR}. 
The model predictions are also shown. The observations suggest a trend that goes in the same direction as the 
predictions of our model.
It would be possible to confirm such a trend if SFRs
were available for all
ETGs in the ATLAS$^{\rm 3D}$, as is the case of the cold gas fraction. 

\section{Limitations of the model and discussion}\label{limitations} 

An important assumption of the model of \citet{Padilla14} is that flips in the angular momenta of the galaxy components are uncorrelated over time. 
This means that an angular momentum flip at a given time, for instance, 
of the gas disk, has a direction uncorrelated to the direction of previous flips.
This is not necessarily the case in reality, as matter can come in from preferred directions, for example from filaments. These filaments 
bring large specific angular momentum compared to that of the halo, and hence are capable of changing the direction of 
the galaxy angular momentum towards the preferred direction of filaments (\citealt{Pichon11}; \citealt{Sales12}; \citealt{Danovich12}; 
\citealt{Aumer13}; \citealt{Danovich14}).
This means that subsequent accretion episodes can come from the same (or similar) directions, which translates into 
angular momentum flips being correlated in direction over time.

\begin{figure}
\begin{center}
\includegraphics[trim = 0mm 0.3mm 1mm 0.45mm,clip,width=0.49\textwidth]{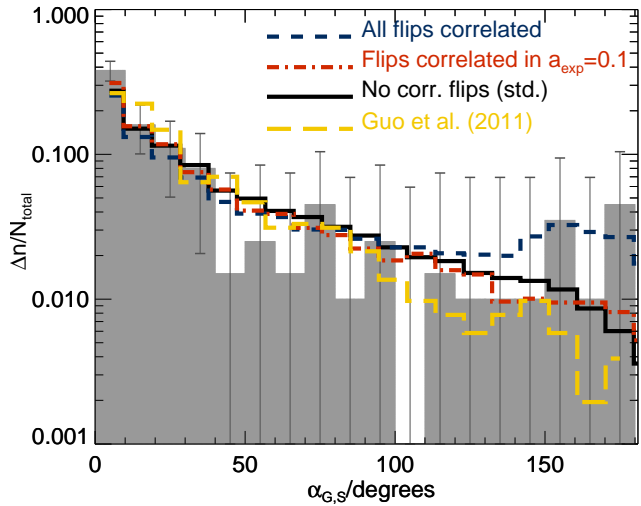}
\caption{Distribution of angles between the angular momenta of the cold gas disk (HI+H$_2$)
and the stellar body, $\alpha_{\rm G,S}$,
for ETGs with $L_{\rm K}>6\times 10^9,L_{\odot}$ and
$M_{\rm HI}+M_{\rm H_2}>10^7\,M_{\odot}$ in the
Lacey14+GRP model, when we consider no correlated angular momentum flips over time 
(standard model; solid line), when all flips are correlated in direction over time (dashed line), 
and when flips are correlated only during finite time intervals (i.e. during an interval of expansion factor $a_{\rm exp}=0.1$; 
dot-dashed line).
We also show the predicted distribution of $\alpha_{\rm G,S}$ from the semi-analytic model of \citet{Guo11}.}
\label{PerVariations}
\end{center}
\end{figure}

\begin{table}
\begin{center}
\caption{The two-sided KS probability, $p_{\rm KS}$, calculated using galaxies from ATLAS$^{\rm 3D}$ and 
the Lacey14+GRP model under the scenario of no correlated flips over time, all flips correlated over time
and flips correlated only over an interval in expansion factor of $a_{\rm exp}=0.1$. We also show the value of 
 $p_{\rm KS}$ for  the publicly available model of \citet{Guo11}.}\label{KStests}
\begin{tabular}{l c }
\\[3pt]
\hline
Model & $p_{\rm KS}$\\
\hline
No flips correlated (Lacey14+GRP)& $0.16$\\
All flips correlated (Lacey14+GRP)& $0.002$\\
Flips correlated over & $0.04$\\
$a_{\rm exp}=0.1$ (Lacey14+GRP) & \\
\hline
\citet{Guo11} & $0.0001$\\
\hline
\end{tabular}
\end{center}
\end{table}

The ability of these cold flows to reach the halo centre where galaxies live is under debate. 
For example, \citet{Nelson13} suggest
that cold streams do not self-shield as efficiently as previously thought, as there is important heating as cold streams
travel through the dark matter halo towards the centre.
They show that the maximum fraction of gas accreted onto galaxies that did not get shock heated to a temperature close to the
virial temperature of the halo is $\approx 0.3$ for a present-day halo mass of $5\times 10^{10}\,M_{\odot}\,h^{-1}$. This fraction decreases
with increasing halo mass. Thus, it is still unclear to what extent angular momentum flips can be correlated in direction over time.

Here we test the idea of all flips in angular momentum being correlated in direction over time, and compare 
this assumption with our standard model, which assumes uncorrelated flips. 
Fig.~\ref{PerVariations} shows the histogram of $\alpha_{\rm G,S}$ for our standard model 
and for the model when all flips are correlated in direction over time.
The latter model produces a flatter distribution of  $\alpha_{\rm G,S}$, particularly at 
$\alpha_{\rm G,S}>100$~degrees. The fraction of galaxies with $\alpha_{\rm G,S}>30$~degrees rises from $46$ per cent in our 
standard model to $52$ per cent. 
In order to quantify which model produces results closer to the observed one 
we calculate the two-sided Kolmogorov-Smirnov (KS) probability, $p_{\rm KS}$ (calculated using the ATLAS$^{\rm 3D}$
results). The values of $p_{\rm KS}$ are shown in Table~\ref{KStests} for the standard model (i.e. uncorrelated flips) and for the realisation 
that assumes correlated flips over time. 

From the values in Table~\ref{KStests}, one can conclude that observations prefer flips in angular momenta that are not  
fully correlated over time, i.e. the direction of flips in angular momentum at any time cannot know about all previous flips. 
Another possibility is that if correlated flips exit, the correlation spans 
timescales that are shorter than the age of galaxies. In order to test this idea, we
 perform an experiment where flips are correlated in direction during a finite period of time. We choose 
to quantify this by the interval in expansion factor where flips are correlated, $a_{\rm exp}$. 
We show the result of this experiment for $a_{\rm exp}=0.1$ in Fig.~\ref{PerVariations} and the results of the 
KS test in Table~\ref{KStests}. This model predicts a fraction of galaxies with $\alpha_{\rm G,S}>30$~degrees 
of $48$ per cent and from the results of Table~\ref{KStests} one can conclude that the model using uncorrelated flips 
is still in better agreement with the observations.

We also show for reference in Fig.~\ref{PerVariations} 
the predictions from a different semi-analytic model, that of \citet{Guo11}. In this model they follow the 3-dimensional angular 
momenta of the gas and stars in galaxies along similar lines to those presented in $\S$~\ref{FlipsModel}. In Guo et al. 
stars and gas are treated as independent bodies and they do not react to the gravity of each other as in 
{\tt GALFORM}. The predicted distribution of $\alpha_{\rm G,S}$ also agrees well with the observations. We calculate 
a KS probability and the result is presented in Table~\ref{KStests}. The Guo et al. model 
gives slighlty worse agreement with the observations of the 
ATLAS$^{\rm 3D}$ than our standard model. The fact that the model 
of Guo et al. following a similar scheme as we presented here also predicts a distribution of 
$\alpha_{\rm G,S}$ in reasonable agreement with the observations points to the robustness of such prediction 
even when physical processes in galaxies are treated differently in different models. 
Unfortunately semi-analytic models are limited by their simplified geometries, and 
it is not possible to include collimated flows feeding galaxies. 
Further studies with fully consistent cosmological 
hydro-dynamic simulations are needed for this task. Cosmological simulations such as {\tt EAGLE} \citep{Schaye14} and 
 {\tt Illustris} \citep{Vogelsberger14} are ideal for this. 

Another interesting result from hydro-dynamical simulations is that the rotation of stars 
can enhance cooling from the hot halo \citep{Negri14}. This may be of particular interest for the fast rotators 
found in ATLAS$^{\rm 3D}$ \citep{Cappellari11b}. However, in the simulation setup of \citet{Negri14}, mass loss from stars 
feeds the halo of galaxies from which cooling takes place. This means that the system is designed to have components which are 
preferentially 
aligned. We show in paper I that the fraction of ETGs have their ISM content dominated by 
this gas source is very small ($\approx 2$ per cent), and therefore this may be a second order effect in our calculations. 
 A more complete study, for example using cosmological hydro-dynamical simulations can constrain the importance of 
this mechanism in driving accretion onto galaxies.

\subsection{Expectations for late-type galaxies}

\begin{figure}
\begin{center}
\includegraphics[trim = 0mm 0.3mm 1mm 0.45mm,clip,width=0.45\textwidth]{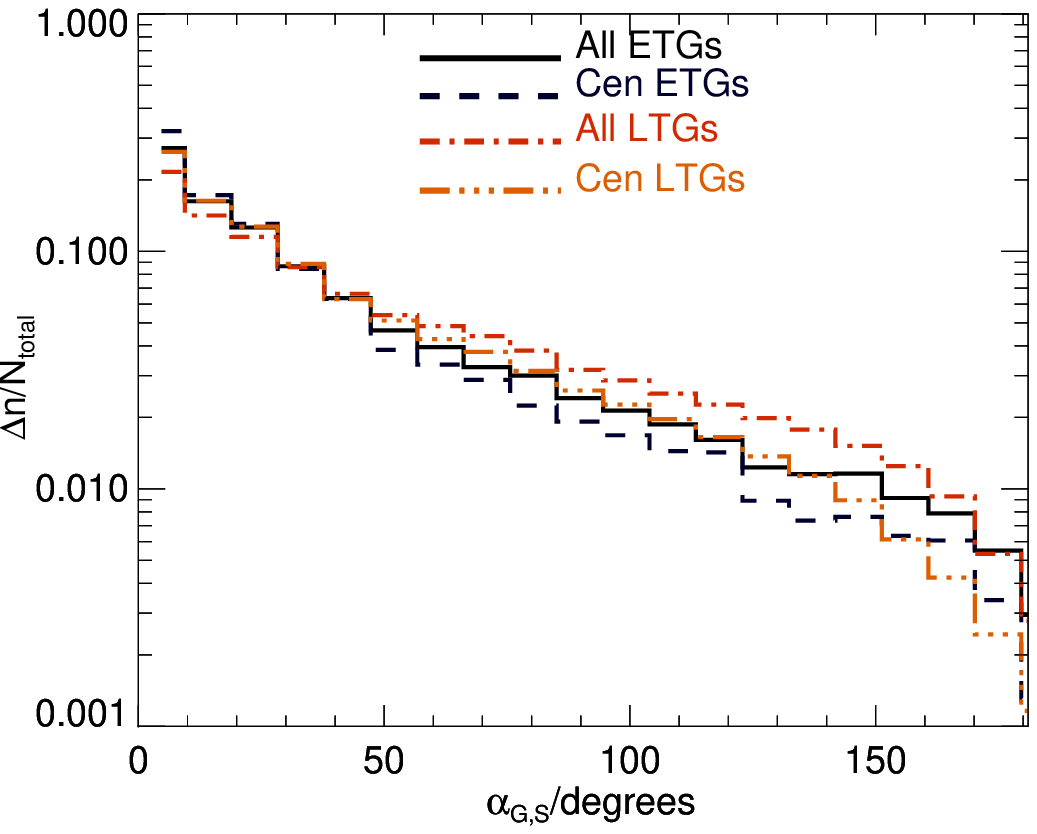}
\includegraphics[trim = 0mm 0.3mm 1mm 0.45mm,clip,width=0.45\textwidth]{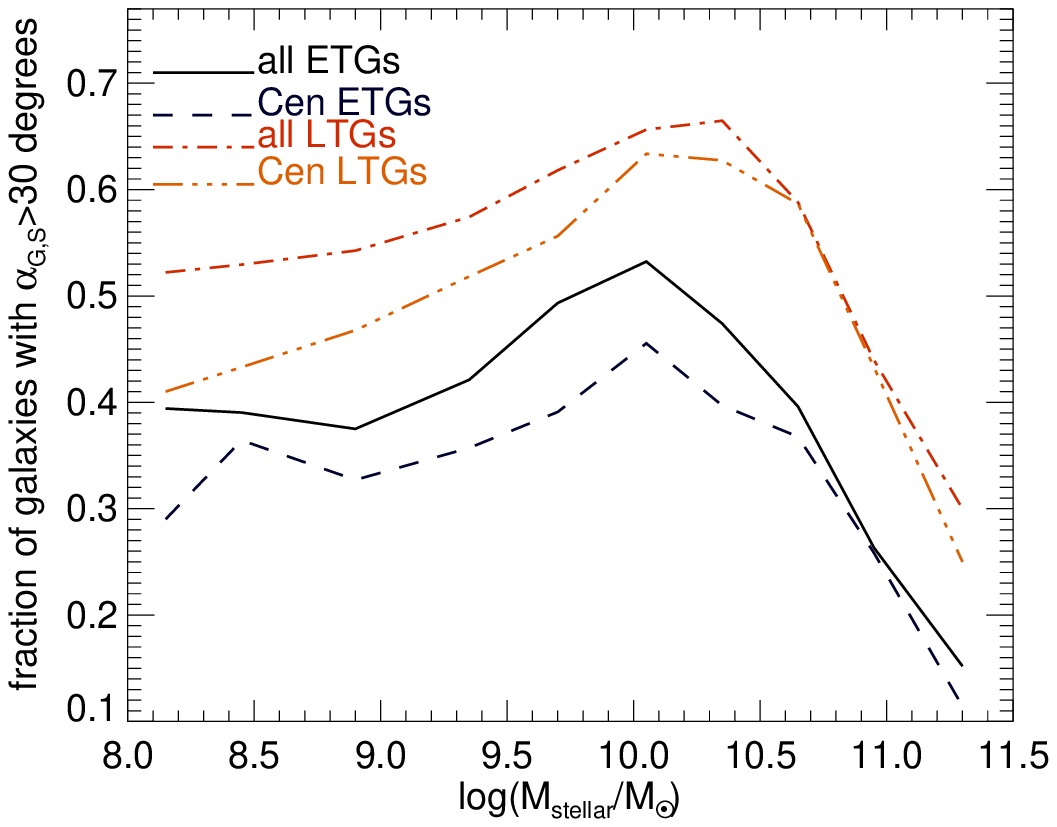}
\caption{{\it Top panel:} Distribution of $\alpha_{\rm G,S}$ for ETGs ($B/T>0.5$) and late-type galaxies (LTGs, $B/T<0.5$) 
with $M_{\rm stellar}>10^8\,M_{\odot}$ and 
$M_{\rm HI}+M_{\rm H_2}>10^7\,M_{\odot}$ in the 
Lacey14+GRP model when we fully apply the \citet{Padilla14} Monte-Carlo method. We show all galaxies of each morphological type, as well 
as centrals only, as labelled. {\it Bottom panel:} The fraction of galaxies with $\alpha_{\rm G,S}>30$~degrees as a function of 
stellar mass for the same model galaxy populations shown in the top panel.}
\label{MisalignmentLTGs}
\end{center}
\end{figure}

The distribution of $\alpha_{\rm G,S}$ for late-type galaxies (LTGs; bulge-to-total stellar mass ratio $<0.5$)
will soon be probed by integral field surveys, such as SAMI, MANGA and CALIFA, and therefore 
it is important to show predictions for this population. The top panel of Fig.~\ref{MisalignmentLTGs} shows the distribution of 
$\alpha_{\rm G,S}$ for LTGs and ETGs that have $M_{\rm stellar}>10^8\,M_{\odot}$ and $M_{\rm HI}+M_{\rm H_2}>10^7\,M_{\odot}$. 
The distribution for central galaxies only is also shown. LTGs are expected to show slightly higher levels of misalignments 
than ETGs. However, the exact fraction of galaxies with $\alpha_{\rm G,S}>30$~degrees is a strong function of 
stellar mass. Note that the anti-correlation discussed in 
$\S$~\ref{DependencyProps} between the fraction of ETGs with $\alpha_{\rm G,S}>30$~degrees and $K$-band luminosity (or stellar mass) reverses 
at $M_{\rm stellar}\approx 10^{10}\,M_{\odot}$ for ETGs and at $M_{\rm stellar}\approx 3\times 10^{10}\,M_{\odot}$ for LTGs, 
with lower mass galaxies having a lower frequency of $\alpha_{\rm G,S}>30$~degrees. 
If this turnover exists, the large galaxy catalogue probed by SAMI will be able to see it. The driver of this turnover is due to  
an important fraction of galaxies below the turnover mass living in haloes with masses 
$M_{\rm subhalo}<5\times 10^{11}\,M_{\odot}$. These galaxies have had a
 quieter history, with fewer interactions relative to galaxies living in more massive haloes. The galaxies in low mass halo 
show more alignments than galaxies in intermediate halo masses (see Fig.~\ref{LBTdoubling}).
 
In the case of LTGs, torques would be stronger since the gas has a clear disk to which to relax. Since we do not explicitly account 
for torques our calculation would be upper limits.

\section{Conclusions}\label{conclusion} 

We have explored the origin of misalignments between the stars and gas in ETGs in the local Universe 
using the Lacey14 variant of the {\tt GALFORM} model. 
The reference observational dataset is the 
ATLAS$^{\rm 3D}$ survey, which displays   
a fraction of 
$42\pm 6$ per cent of ETGs exhibiting 
angles between the angular momenta of the gas and stars of 
 $\alpha_{\rm G,S}>30$~degrees.
This paper is the second of the series. We show in 
paper I \citep{Lagos14} that the Lacey14 model reproduces the HI and H$_2$ gas fractions and gas mass distribution functions 
of ETGs very well. The key to the latter 
is the incorporation of gradual ram pressure stripping (see \citealt{Lagos14} for details). 

We first explore the simplest assumption, which is that the only source 
of misaligned cold gas (HI plus H$_2$) with respect to the stellar body 
in ETGs is galaxy mergers. This has traditionally been the observational interpretation of 
 of misaligned cold gas disks (\citealt{Davis11}; \citet{Serra12}).  However, we find that 
there are not enough galaxy mergers at $z\approx 0$ in the model, regardless of the 
dynamical friction timescale adopted. We find that by adopting the \citet{Jiang07} dynamical timescale, which is the default option 
 in Lacey14, only $11$ per cent of ETGs in the model have had a minor merger in the last $1$~Gyr that involved important 
gas accretion, leading to only $2$ per cent of ETGs having $\alpha_{\rm G,S}>30$~degrees. This percentage 
rises to $\approx 5$ per cent if we use the \citet{Lacey93} dynamical friction timescale, which favours more recent minor mergers. 

We then explore a more general approach, which is to follow the history of gas accretion and stellar growth due to star formation 
in-situ and due to galaxy mergers, and apply the Monte-Carlo simulation of \citet{Padilla14} to follow the angular momenta flips 
that are produced by this growth. This simulation analyses the change in mass and the source of that change (i.e. smooth accretion or 
mergers) and returns PDFs of the angle between the direction of the angular momentum of the component that is changing in mass before 
and after the accretion. That angle can change in any direction and we accumulate the change over time. We do this for all the 
components of ETGs we are interested in: hot halo, cold gas disk (HI plus H$_2$) and stellar component. We apply 
this procedure for all ETGs in the semi-analytic model that have been selected to mimic the ETG selection in the 
ATLAS$^{\rm 3D}$ survey (based on their $K$-band luminosity and HI+H$_2$ mass). Our main findings are:

\begin{itemize}
\item By applying the Padilla et al. Monte-Carlo simulation to the ETGs predicted by the Lacey14 model, we find that  
$46$ per cent of modelled ETGs have $\alpha_{\rm G,S}>30$~degrees. This angle varies in the range $35-53$ per cent when we select sub-volumes 
of the simulation of the size 
of the volume probed by the ATLAS$^{\rm 3D}$ (which observed all ETGs with $L_{\rm K}>6\times 10^{9}L_{\odot}$ 
out to $42$~Mpc). The level of misalignment found agrees with the observations, within the error bars.
Note that our model has not been tuned to reproduce this observable and therefore this is a true prediction of the model.
\item We find that misalignments ($\alpha_{\rm G,S}>30$~degrees) are found in ETGs living in 
intermediate mass haloes ($7\times 10^{10}\,M_{\odot}<M_{\rm halo}/h<
10^{13}\,M_{\odot}$), while ETGs hosted by larger or lower mass haloes tend to have $\alpha_{\rm G,S}<30$~degrees. This agrees with the 
 observational results that ETGs living very close to each other (e.g. galaxy clusters or massive galaxy groups) show stellar and gas components 
that are aligned \citep{Davis11}. 
\item We find that the relation between $\alpha_{\rm G,S}$ and the halo mass leads to correlations between 
$\alpha_{\rm G,S}$, stellar mass, cold gas fraction and SFR. Galaxies in the model with 
high stellar masses, low SFRs and low cold gas fractions show lower values of $\alpha_{\rm G,S}$.
The observational findings from ATLAS$^{\rm 3D}$ shows that galaxies with $L_{\rm K}\gtrsim 3\times 10^{11}L_{\odot}$
show $\alpha_{\rm G,S}<30$~degrees, which agrees very well with our predictions. These show that 
massive galaxies, $L_{\rm K}>3\times 10^{11}\,L_{\odot}$, are very unlikely to show misalignments. 
We also test the prediction that there is a dependence of $\alpha_{\rm G,S}$ on cold gas fraction and SFR
using the observations of the ATLAS$^{\rm 3D}$ and we find that these correlations are also in the observations, and they had not 
been found before.
\item The above trends are understood within the simple phenomenological finding that {\it galaxies that have important gas accretion  
after most of the stellar mass of the galaxy is in place, are more likely to exhibit cold gas disks misaligned with respect to the 
stars}. This accreted gas can come from either
galaxy mergers or smooth cold gas accretion and it is not necessarily aligned with the existing stellar body. 
Galaxies in the model living in galaxy groups and clusters, 
which are also the most massive ETGs, lack important recent gas accretion 
episodes, and therefore their cold gas and stellar mass are aligned.
\item We show expectations for LTGs and ETGs over a wide range of stellar masses ($M_{\rm stellar}>10^8\,M_{\odot}$) and predict the existence 
of a turnover mass ($M_{\rm to}\approx 10^{10}\,M_{\odot}$), at which misalignments are most likely
The existence of this turnover mass will soon be probed by 
integral field galaxy surveys.
\end{itemize}
 
The Monte-Carlo approach applied to {\tt GALFORM} galaxies offers new, interesting ideas for the origin of what has been observed 
in ETGs in the local Universe, which has proven elusive in simulations of individual galaxies \citep{Serra14}. 
The next step will be to integrate this angular momentum scheme directly into {\tt GALFORM} to model intrinsic alignments between 
galaxies. Large 
improvements can also come from recent hydro-dynamical simulations \citep{Vogelsberger14,Schaye14}, which for the first time 
can tackle the problem of the angular momenta of the different galaxy components from 
hydro-dynamical simulations in a statistical fashion.

\section*{Acknowledgements}

The authors thank Andreas Schruba, Paolo Serra, Qi Guo, Lisa Young and Madhura Killedar for useful discussions. 
The research leading to these results has received funding from 
the European Community's Seventh
Framework Programme ($/$FP7$/$2007-2013$/$) under grant agreement no 229517 and 
the STFC consolidated grant ST/L00075X/I.
VGP acknowledges support from a European Research
Council Starting Grant (DEGAS-259586).
This work used the DiRAC Data Centric system at Durham University, operated by the Institute for Computational Cosmology on behalf of the STFC DiRAC HPC Facility ({\tt www.dirac.ac.uk}). This equipment was funded by BIS National E-infrastructure capital grant ST/K00042X/1, STFC capital grant ST/H008519/1, and STFC DiRAC Operations grant ST/K003267/1 and Durham University. DiRAC is part of the National E-Infrastructure.

\bibliographystyle{mn2e_trunc8}
\bibliography{OriginGasContentEllsII}

\begin{thebibliography}{86}
\expandafter\ifx\csname natexlab\endcsname\relax\def\natexlab#1{#1}\fi

\bibitem[{{Alatalo} {et~al.}(2013){Alatalo}, {Davis}, {Bureau}, {Young},
  {Blitz}, {Crocker}, {Bayet}, {Bois}, {Bournaud}, {Cappellari}, {Davies}, {de
  Zeeuw}, {Duc}, {Emsellem}, {Khochfar}, {Krajnovi{\'c}}, {Kuntschner},
  {Lablanche}, {Morganti}, {McDermid}, {Naab}, {Oosterloo}, {Sarzi}, {Scott},
  {Serra}, \& {Weijmans}}]{Alatalo13}
{Alatalo} K., {Davis} T.~A., {Bureau} M., {Young} L.~M., {Blitz} L., {Crocker}
  A.~F., {Bayet} E., {Bois} M. {et~al}, 2013, \mnras, 432, 1796

\bibitem[{{Aumer} {et~al.}(2013){Aumer}, {White}, {Naab}, \&
  {Scannapieco}}]{Aumer13}
{Aumer} M., {White} S.~D.~M., {Naab} T., {Scannapieco} C., 2013, \mnras, 434,
  3142

\bibitem[{{Baugh} {et~al.}(1996){Baugh}, {Cole}, \& {Frenk}}]{Baugh96}
{Baugh} C.~M., {Cole} S., {Frenk} C.~S., 1996, \mnras, 283, 1361

\bibitem[{{Baugh} {et~al.}(2005){Baugh}, {Lacey}, {Frenk}, {Granato}, {Silva},
  {Bressan}, {Benson}, \& {Cole}}]{Baugh05}
{Baugh} C.~M., {Lacey} C.~G., {Frenk} C.~S., {Granato} G.~L., {Silva} L.,
  {Bressan} A., {Benson} A.~J., {Cole} S., 2005, \mnras, 356, 1191

\bibitem[{{Benson} \& {Bower}(2010)}]{Benson10}
{Benson} A.~J., {Bower} R., 2010, \mnras, 405, 1573

\bibitem[{{Benson} {et~al.}(2002){Benson}, {Frenk}, {Lacey}, {Baugh}, \&
  {Cole}}]{Benson02}
{Benson} A.~J., {Frenk} C.~S., {Lacey} C.~G., {Baugh} C.~M., {Cole} S., 2002,
  \mnras, 333, 177

\bibitem[{{Bett} {et~al.}(2010){Bett}, {Eke}, {Frenk}, {Jenkins}, \&
  {Okamoto}}]{Bett10}
{Bett} P., {Eke} V., {Frenk} C.~S., {Jenkins} A., {Okamoto} T., 2010, \mnras,
  404, 1137

\bibitem[{{Bett} \& {Frenk}(2012)}]{Bett12}
{Bett} P.~E., {Frenk} C.~S., 2012, \mnras, 420, 3324

\bibitem[{{Binney} \& {May}(1986)}]{Binney86}
{Binney} J., {May} A., 1986, \mnras, 218, 743

\bibitem[{{Blitz} \& {Rosolowsky}(2006)}]{Blitz06}
{Blitz} L., {Rosolowsky} E., 2006, \apj, 650, 933

\bibitem[{{Boselli} {et~al.}(2014){Boselli}, {Cortese}, {Boquien}, {Boissier},
  {Catinella}, {Gavazzi}, {Lagos}, \& {Saintonge}}]{Boselli14}
{Boselli} A., {Cortese} L., {Boquien} M., {Boissier} S., {Catinella} B.,
  {Gavazzi} G., {Lagos} C., {Saintonge} A., 2014, ArXiv e-prints

\bibitem[{{Bournaud} {et~al.}(2011){Bournaud}, {Chapon}, {Teyssier}, {Powell},
  {Elmegreen}, {Elmegreen}, {Duc}, {Contini}, {Epinat}, \&
  {Shapiro}}]{Bournaud11}
{Bournaud} F., {Chapon} D., {Teyssier} R., {Powell} L.~C., {Elmegreen} B.~G.,
  {Elmegreen} D.~M., {Duc} P.-A., {Contini} T. {et~al}, 2011, \apj, 730, 4

\bibitem[{{Bournaud} {et~al.}(2014){Bournaud}, {Perret}, {Renaud}, {Dekel},
  {Elmegreen}, {Elmegreen}, {Teyssier}, {Amram}, {Daddi}, {Duc}, {Elbaz},
  {Epinat}, {Gabor}, {Juneau}, {Kraljic}, \& {Le Floch'}}]{Bournaud14}
{Bournaud} F., {Perret} V., {Renaud} F., {Dekel} A., {Elmegreen} B.~G.,
  {Elmegreen} D.~M., {Teyssier} R., {Amram} P. {et~al}, 2014, \apj, 780, 57

\bibitem[{{Bower} {et~al.}(2006){Bower}, {Benson}, {Malbon}, {Helly}, {Frenk},
  {Baugh}, {Cole}, \& {Lacey}}]{Bower06}
{Bower} R.~G., {Benson} A.~J., {Malbon} R., {Helly} J.~C., {Frenk} C.~S.,
  {Baugh} C.~M., {Cole} S., {Lacey} C.~G., 2006, \mnras, 370, 645

\bibitem[{{Boylan-Kolchin} {et~al.}(2009){Boylan-Kolchin}, {Springel}, {White},
  {Jenkins}, \& {Lemson}}]{Boylan-Kolchin09}
{Boylan-Kolchin} M., {Springel} V., {White} S.~D.~M., {Jenkins} A., {Lemson}
  G., 2009, \mnras, 398, 1150

\bibitem[{{Bryan} {et~al.}(2013){Bryan}, {Kay}, {Duffy}, {Schaye}, {Dalla
  Vecchia}, \& {Booth}}]{Bryan13}
{Bryan} S.~E., {Kay} S.~T., {Duffy} A.~R., {Schaye} J., {Dalla Vecchia} C.,
  {Booth} C.~M., 2013, \mnras, 429, 3316

\bibitem[{{Cappellari} {et~al.}(2011){Cappellari}, {Emsellem}, {Krajnovi{\'c}},
  {McDermid}, {Serra}, {Alatalo}, {Blitz}, {Bois}, {Bournaud}, {Bureau},
  {Davies}, {Davis}, {de Zeeuw}, {Khochfar}, {Kuntschner}, {Lablanche},
  {Morganti}, {Naab}, {Oosterloo}, {Sarzi}, {Scott}, {Weijmans}, \&
  {Young}}]{Cappellari11b}
{Cappellari} M., {Emsellem} E., {Krajnovi{\'c}} D., {McDermid} R.~M., {Serra}
  P., {Alatalo} K., {Blitz} L., {Bois} M. {et~al}, 2011, \mnras, 416, 1680

\bibitem[{{Cavaliere} \& {Fusco-Femiano}(1976)}]{Cavaliere76}
{Cavaliere} A., {Fusco-Femiano} R., 1976, \aap, 49, 137

\bibitem[{{Ceverino} {et~al.}(2010){Ceverino}, {Dekel}, \&
  {Bournaud}}]{Ceverino10}
{Ceverino} D., {Dekel} A., {Bournaud} F., 2010, \mnras, 404, 2151

\bibitem[{{Cole} {et~al.}(2000){Cole}, {Lacey}, {Baugh}, \& {Frenk}}]{Cole00}
{Cole} S., {Lacey} C.~G., {Baugh} C.~M., {Frenk} C.~S., 2000, \mnras, 319, 168

\bibitem[{{Cortese} {et~al.}(2011){Cortese}, {Catinella}, {Boissier},
  {Boselli}, \& {Heinis}}]{Cortese11}
{Cortese} L., {Catinella} B., {Boissier} S., {Boselli} A., {Heinis} S., 2011,
  \mnras, 415, 1797

\bibitem[{{Creasey} {et~al.}(2013){Creasey}, {Theuns}, \& {Bower}}]{Creasey12}
{Creasey} P., {Theuns} T., {Bower} R.~G., 2013, \mnras, 429, 1922

\bibitem[{{Croom} {et~al.}(2012){Croom}, {Lawrence}, {Bland-Hawthorn},
  {Bryant}, {Fogarty}, {Richards}, {Goodwin}, {Farrell}, {Miziarski}, {Heald},
  {Jones}, {Lee}, {Colless}, {Brough}, {Hopkins}, {Bauer}, {Birchall}, {Ellis},
  {Horton}, {Leon-Saval}, {Lewis}, {L{\'o}pez-S{\'a}nchez}, {Min}, {Trinh}, \&
  {Trowland}}]{Croom12}
{Croom} S.~M., {Lawrence} J.~S., {Bland-Hawthorn} J., {Bryant} J.~J., {Fogarty}
  L., {Richards} S., {Goodwin} M., {Farrell} T. {et~al}, 2012, \mnras, 421, 872

\bibitem[{{Danovich} {et~al.}(2014){Danovich}, {Dekel}, {Hahn}, {Ceverino}, \&
  {Primack}}]{Danovich14}
{Danovich} M., {Dekel} A., {Hahn} O., {Ceverino} D., {Primack} J., 2014,
  ArXiv:1407.7129

\bibitem[{{Danovich} {et~al.}(2012){Danovich}, {Dekel}, {Hahn}, \&
  {Teyssier}}]{Danovich12}
{Danovich} M., {Dekel} A., {Hahn} O., {Teyssier} R., 2012, \mnras, 422, 1732

\bibitem[{{Davis} {et~al.}(1985){Davis}, {Efstathiou}, {Frenk}, \&
  {White}}]{Davis85}
{Davis} M., {Efstathiou} G., {Frenk} C.~S., {White} S.~D.~M., 1985, \apj, 292,
  371

\bibitem[{{Davis} {et~al.}(2013){Davis}, {Alatalo}, {Bureau}, {Cappellari},
  {Scott}, {Young}, {Blitz}, {Crocker}, {Bayet}, {Bois}, {Bournaud}, {Davies},
  {de Zeeuw}, {Duc}, {Emsellem}, {Khochfar}, {Krajnovi{\'c}}, {Kuntschner},
  {Lablanche}, {McDermid}, {Morganti}, {Naab}, {Oosterloo}, {Sarzi}, {Serra},
  \& {Weijmans}}]{Davis13}
{Davis} T.~A., {Alatalo} K., {Bureau} M., {Cappellari} M., {Scott} N., {Young}
  L.~M., {Blitz} L., {Crocker} A. {et~al}, 2013, \mnras, 429, 534

\bibitem[{{Davis} {et~al.}(2011){Davis}, {Bureau}, {Young}, {Alatalo}, {Blitz},
  {Cappellari}, {Scott}, {Bois}, {Bournaud}, {Davies}, {de Zeeuw}, {Emsellem},
  {Khochfar}, {Krajnovi{\'c}}, {Kuntschner}, {Lablanche}, {McDermid},
  {Morganti}, {Naab}, {Oosterloo}, {Sarzi}, {Serra}, \& {Weijmans}}]{Davis11}
{Davis} T.~A., {Bureau} M., {Young} L.~M., {Alatalo} K., {Blitz} L.,
  {Cappellari} M., {Scott} N., {Bois} M. {et~al}, 2011, \mnras, 414, 968

\bibitem[{{Davis} {et~al.}(2014){Davis}, {Young}, {Crocker}, {Bureau}, {Blitz},
  {Alatalo}, {Emsellem}, {Naab}, {Bayet}, {Bois}, {Bournaud}, {Cappellari},
  {Davies}, {de Zeeuw}, {Duc}, {Khochfar}, {Krajnovic}, {Kuntschner},
  {McDermid}, {Morganti}, {Oosterloo}, {Sarzi}, {Scott}, {Serra}, \&
  {Weijmans}}]{Davis14}
{Davis} T.~A., {Young} L.~M., {Crocker} A.~F., {Bureau} M., {Blitz} L.,
  {Alatalo} K., {Emsellem} E., {Naab} T. {et~al}, 2014, ArXiv:1403.4850

\bibitem[{{Dekel} {et~al.}(2009){Dekel}, {Sari}, \& {Ceverino}}]{Dekel09}
{Dekel} A., {Sari} R., {Ceverino} D., 2009, \apj, 703, 785

\bibitem[{{Driver} \& {Robotham}(2010)}]{Driver10}
{Driver} S.~P., {Robotham} A.~S.~G., 2010, \mnras, 407, 2131

\bibitem[{{Dubois} {et~al.}(2014){Dubois}, {Pichon}, {Welker}, {Le Borgne},
  {Devriendt}, {Laigle}, {Codis}, {Pogosyan}, {Arnouts}, {Benabed}, {Bertin},
  {Blaizot}, {Bouchet}, {Cardoso}, {Colombi}, {de Lapparent}, {Desjacques},
  {Gavazzi}, {Kassin}, {Kimm}, {McCracken}, {Milliard}, {Peirani}, {Prunet},
  {Rouberol}, {Silk}, {Slyz}, {Sousbie}, {Teyssier}, {Tresse}, {Treyer},
  {Vibert}, \& {Volonteri}}]{Dubois14}
{Dubois} Y., {Pichon} C., {Welker} C., {Le Borgne} D., {Devriendt} J., {Laigle}
  C., {Codis} S., {Pogosyan} D. {et~al}, 2014, \mnras, 444, 1453

\bibitem[{{Efstathiou} {et~al.}(1982){Efstathiou}, {Lake}, \&
  {Negroponte}}]{Efstathiou82}
{Efstathiou} G., {Lake} G., {Negroponte} J., 1982, \mnras, 199, 1069

\bibitem[{{Emsellem} {et~al.}(2004){Emsellem}, {Cappellari}, {Peletier},
  {McDermid}, {Bacon}, {Bureau}, {Copin}, {Davies}, {Krajnovi{\'c}},
  {Kuntschner}, {Miller}, \& {de Zeeuw}}]{Emsellem04}
{Emsellem} E., {Cappellari} M., {Peletier} R.~F., {McDermid} R.~M., {Bacon} R.,
  {Bureau} M., {Copin} Y., {Davies} R.~L. {et~al}, 2004, \mnras, 352, 721

\bibitem[{{Fanidakis} {et~al.}(2012){Fanidakis}, {Baugh}, {Benson}, {Bower},
  {Cole}, {Done}, {Frenk}, {Hickox}, {Lacey}, \& {Del P.~Lagos}}]{Fanidakis10b}
{Fanidakis} N., {Baugh} C.~M., {Benson} A.~J., {Bower} R.~G., {Cole} S., {Done}
  C., {Frenk} C.~S., {Hickox} R.~C. {et~al}, 2012, \mnras, 419, 2797

\bibitem[{{Font} {et~al.}(2008){Font}, {Bower}, {McCarthy}, {Benson}, {Frenk},
  {Helly}, {Lacey}, {Baugh}, \& {Cole}}]{Font08}
{Font} A.~S., {Bower} R.~G., {McCarthy} I.~G., {Benson} A.~J., {Frenk} C.~S.,
  {Helly} J.~C., {Lacey} C.~G., {Baugh} C.~M. {et~al}, 2008, \mnras, 389, 1619

\bibitem[{{Gallagher} {et~al.}(1975){Gallagher}, {Faber}, \&
  {Balick}}]{Gallagher75}
{Gallagher} J.~S., {Faber} S.~M., {Balick} B., 1975, \apj, 202, 7

\bibitem[{{Gonzalez-Perez} {et~al.}(2014){Gonzalez-Perez}, {Lacey}, {Baugh},
  {Lagos}, {Helly}, {Campbell}, \& {Mitchell}}]{Gonzalez-Perez13}
{Gonzalez-Perez} V., {Lacey} C.~G., {Baugh} C.~M., {Lagos} C.~D.~P., {Helly}
  J., {Campbell} D.~J.~R., {Mitchell} P.~D., 2014, \mnras

\bibitem[{{Guo} {et~al.}(2011){Guo}, {White}, {Boylan-Kolchin}, {De Lucia},
  {Kauffmann}, {Lemson}, {Li}, {Springel}, \& {Weinmann}}]{Guo11}
{Guo} Q., {White} S., {Boylan-Kolchin} M., {De Lucia} G., {Kauffmann} G.,
  {Lemson} G., {Li} C., {Springel} V. {et~al}, 2011, \mnras, 413, 101

\bibitem[{{Husemann} {et~al.}(2013){Husemann}, {Jahnke}, {S{\'a}nchez},
  {Barrado}, {Bekerait*error*{\.e}}, {Bomans}, {Castillo-Morales},
  {Catal{\'a}n-Torrecilla}, {Cid Fernandes}, {Falc{\'o}n-Barroso},
  {Garc{\'{\i}}a-Benito}, {Gonz{\'a}lez Delgado}, {Iglesias-P{\'a}ramo},
  {Johnson}, {Kupko}, {L{\'o}pez-Fernandez}, {Lyubenova}, {Marino}, {Mast},
  {Miskolczi}, {Monreal-Ibero}, {Gil de Paz}, {P{\'e}rez}, {P{\'e}rez},
  {Rosales-Ortega}, {Ruiz-Lara}, {Schilling}, {van de Ven}, {Walcher}, {Alves},
  {de Amorim}, {Backsmann}, {Barrera-Ballesteros}, {Bland-Hawthorn}, {Cortijo},
  {Dettmar}, {Demleitner}, {D{\'{\i}}az}, {Enke}, {Florido}, {Flores},
  {Galbany}, {Gallazzi}, {Garc{\'{\i}}a-Lorenzo}, {Gomes}, {Gruel}, {Haines},
  {Holmes}, {Jungwiert}, {Kalinova}, {Kehrig}, {Kennicutt}, {Klar}, {Lehnert},
  {L{\'o}pez-S{\'a}nchez}, {de Lorenzo-C{\'a}ceres}, {M{\'a}rmol-Queralt{\'o}},
  {M{\'a}rquez}, {Mendez-Abreu}, {Moll{\'a}}, {del Olmo}, {Meidt}, {Papaderos},
  {Puschnig}, {Quirrenbach}, {Roth}, {S{\'a}nchez-Bl{\'a}zquez}, {Spekkens},
  {Singh}, {Stanishev}, {Trager}, {Vilchez}, {Wild}, {Wisotzki}, {Zibetti}, \&
  {Ziegler}}]{Husemann13}
{Husemann} B., {Jahnke} K., {S{\'a}nchez} S.~F., {Barrado} D.,
  {Bekerait*error*{\.e}} S., {Bomans} D.~J., {Castillo-Morales} A.,
  {Catal{\'a}n-Torrecilla} C. {et~al}, 2013, \aap, 549, A87

\bibitem[{{Jiang} {et~al.}(2008){Jiang}, {Jing}, {Faltenbacher}, {Lin}, \&
  {Li}}]{Jiang07}
{Jiang} C.~Y., {Jing} Y.~P., {Faltenbacher} A., {Lin} W.~P., {Li} C., 2008,
  \apj, 675, 1095

\bibitem[{{Jiang} {et~al.}(2014){Jiang}, {Helly}, {Cole}, \& {Frenk}}]{Jiang14}
{Jiang} L., {Helly} J.~C., {Cole} S., {Frenk} C.~S., 2014, \mnras, 440, 2115

\bibitem[{{Jim{\'e}nez} {et~al.}(2011){Jim{\'e}nez}, {Cora}, {Bassino},
  {Tecce}, \& {Smith Castelli}}]{Jimenez11}
{Jim{\'e}nez} N., {Cora} S.~A., {Bassino} L.~P., {Tecce} T.~E., {Smith
  Castelli} A.~V., 2011, \mnras, 417, 785

\bibitem[{{Johansson} {et~al.}(2012){Johansson}, {Naab}, \&
  {Ostriker}}]{Johansson12}
{Johansson} P.~H., {Naab} T., {Ostriker} J.~P., 2012, \apj, 754, 115

\bibitem[{{Kauffmann}(1996)}]{Kauffmann96}
{Kauffmann} G., 1996, \mnras, 281, 487

\bibitem[{{Kennicutt}(1983)}]{Kennicutt83}
{Kennicutt} Jr. R.~C., 1983, \apj, 272, 54

\bibitem[{{Knebe} {et~al.}(2013){Knebe}, {Pearce}, {Lux}, {Ascasibar},
  {Behroozi}, {Casado}, {Moran}, {Diemand}, {Dolag}, {Dominguez-Tenreiro},
  {Elahi}, {Falck}, {Gottl{\"o}ber}, {Han}, {Klypin}, {Luki{\'c}},
  {Maciejewski}, {McBride}, {Merch{\'a}n}, {Muldrew}, {Neyrinck}, {Onions},
  {Planelles}, {Potter}, {Quilis}, {Rasera}, {Ricker}, {Roy}, {Ruiz},
  {Sgr{\'o}}, {Springel}, {Stadel}, {Sutter}, {Tweed}, \& {Zemp}}]{Knebe13}
{Knebe} A., {Pearce} F.~R., {Lux} H., {Ascasibar} Y., {Behroozi} P., {Casado}
  J., {Moran} C.~C., {Diemand} J. {et~al}, 2013, \mnras, 435, 1618

\bibitem[{{Komatsu} {et~al.}(2011){Komatsu}, {Smith}, {Dunkley}, {Bennett},
  {Gold}, {Hinshaw}, {Jarosik}, {Larson}, {Nolta}, {Page}, {Spergel},
  {Halpern}, {Hill}, {Kogut}, {Limon}, {Meyer}, {Odegard}, {Tucker}, {Weiland},
  {Wollack}, \& {Wright}}]{Komatsu11}
{Komatsu} E., {Smith} K.~M., {Dunkley} J., {Bennett} C.~L., {Gold} B.,
  {Hinshaw} G., {Jarosik} N., {Larson} D. {et~al}, 2011, \apjs, 192, 18

\bibitem[{{Lacey} \& {Cole}(1993)}]{Lacey93}
{Lacey} C., {Cole} S., 1993, \mnras, 262, 627

\bibitem[{{Lagos} {et~al.}(2011{\natexlab{a}}){Lagos}, {Baugh}, {Lacey},
  {Benson}, {Kim}, \& {Power}}]{Lagos11}
{Lagos} C.~D.~P., {Baugh} C.~M., {Lacey} C.~G., {Benson} A.~J., {Kim} H.-S.,
  {Power} C., 2011{\natexlab{a}}, \mnras, 418, 1649

\bibitem[{{Lagos} {et~al.}(2014{\natexlab{a}}){Lagos}, {Baugh}, {Zwaan},
  {Lacey}, {Gonzalez-Perez}, {Power}, {Swinbank}, \& {van Kampen}}]{Lagos14b}
{Lagos} C.~D.~P., {Baugh} C.~M., {Zwaan} M.~A., {Lacey} C.~G., {Gonzalez-Perez}
  V., {Power} C., {Swinbank} A.~M., {van Kampen} E., 2014{\natexlab{a}},
  \mnras, 440, 920

\bibitem[{{Lagos} {et~al.}(2012){Lagos}, {Bayet}, {Baugh}, {Lacey}, {Bell},
  {Fanidakis}, \& {Geach}}]{Lagos12}
{Lagos} C.~d.~P., {Bayet} E., {Baugh} C.~M., {Lacey} C.~G., {Bell} T.~A.,
  {Fanidakis} N., {Geach} J.~E., 2012, \mnras, 426, 2142

\bibitem[{{Lagos} {et~al.}(2008){Lagos}, {Cora}, \& {Padilla}}]{Lagos08}
{Lagos} C.~D.~P., {Cora} S.~A., {Padilla} N.~D., 2008, \mnras, 388, 587

\bibitem[{{Lagos} {et~al.}(2014{\natexlab{b}}){Lagos}, {Davis}, {Lacey},
  {Zwaan}, {Baugh}, {Gonzalez-Perez}, \& {Padilla}}]{Lagos14}
{Lagos} C.~d.~P., {Davis} T.~A., {Lacey} C.~G., {Zwaan} M.~A., {Baugh} C.~M.,
  {Gonzalez-Perez} V., {Padilla} N.~D., 2014{\natexlab{b}}, \mnras, 443, 1002

\bibitem[{{Lagos} {et~al.}(2013){Lagos}, {Lacey}, \& {Baugh}}]{Lagos13}
{Lagos} C.~d.~P., {Lacey} C.~G., {Baugh} C.~M., 2013, \mnras, 436, 1787

\bibitem[{{Lagos} {et~al.}(2011{\natexlab{b}}){Lagos}, {Lacey}, {Baugh},
  {Bower}, \& {Benson}}]{Lagos10}
{Lagos} C.~D.~P., {Lacey} C.~G., {Baugh} C.~M., {Bower} R.~G., {Benson} A.~J.,
  2011{\natexlab{b}}, \mnras, 416, 1566

\bibitem[{{Marigo}(2001)}]{Marigo01}
{Marigo} P., 2001, \aap, 370, 194

\bibitem[{{McCarthy} {et~al.}(2008){McCarthy}, {Frenk}, {Font}, {Lacey},
  {Bower}, {Mitchell}, {Balogh}, \& {Theuns}}]{McCarthy08}
{McCarthy} I.~G., {Frenk} C.~S., {Font} A.~S., {Lacey} C.~G., {Bower} R.~G.,
  {Mitchell} N.~L., {Balogh} M.~L., {Theuns} T., 2008, \mnras, 383, 593

\bibitem[{{Mitchell} {et~al.}(2014){Mitchell}, {Lacey}, {Cole}, \&
  {Baugh}}]{Mitchell14}
{Mitchell} P.~D., {Lacey} C.~G., {Cole} S., {Baugh} C.~M., 2014, \mnras, 444,
  2637

\bibitem[{{Mo} {et~al.}(1998){Mo}, {Mao}, \& {White}}]{Mo98}
{Mo} H.~J., {Mao} S., {White} S.~D.~M., 1998, \mnras, 295, 319

\bibitem[{{Morganti} {et~al.}(2006){Morganti}, {de Zeeuw}, {Oosterloo},
  {McDermid}, {Krajnovi{\'c}}, {Cappellari}, {Kenn}, {Weijmans}, \&
  {Sarzi}}]{Morganti06}
{Morganti} R., {de Zeeuw} P.~T., {Oosterloo} T.~A., {McDermid} R.~M.,
  {Krajnovi{\'c}} D., {Cappellari} M., {Kenn} F., {Weijmans} A. {et~al}, 2006,
  \mnras, 371, 157

\bibitem[{{Murray} {et~al.}(2005){Murray}, {Quataert}, \&
  {Thompson}}]{Murray05}
{Murray} N., {Quataert} E., {Thompson} T.~A., 2005, \apj, 618, 569

\bibitem[{{Naab} {et~al.}(2013){Naab}, {Oser}, {Emsellem}, {Cappellari},
  {Krajnovic}, {McDermid}, {Alatalo}, {Bayet}, {Blitz}, {Bois}, {Bournaud},
  {Bureau}, {Crocker}, {Davies}, {Davis}, {de Zeeuw}, {Duc}, {Hirschmann},
  {Johansson}, {Khochfar}, {Kuntschner}, {Morganti}, {Oosterloo}, {Sarzi},
  {Scott}, {Serra}, {van de Ven}, {Weijmans}, \& {Young}}]{Naab13}
{Naab} T., {Oser} L., {Emsellem} E., {Cappellari} M., {Krajnovic} D.,
  {McDermid} R.~M., {Alatalo} K., {Bayet} E. {et~al}, 2013, ArXiv:1311.0284

\bibitem[{{Negri} {et~al.}(2014){Negri}, {Posacki}, {Pellegrini}, \&
  {Ciotti}}]{Negri14}
{Negri} A., {Posacki} S., {Pellegrini} S., {Ciotti} L., 2014, ArXiv:1406.0008

\bibitem[{{Nelson} {et~al.}(2013){Nelson}, {Vogelsberger}, {Genel}, {Sijacki},
  {Kere{\v s}}, {Springel}, \& {Hernquist}}]{Nelson13}
{Nelson} D., {Vogelsberger} M., {Genel} S., {Sijacki} D., {Kere{\v s}} D.,
  {Springel} V., {Hernquist} L., 2013, \mnras, 429, 3353

\bibitem[{{Obreschkow} \& {Glazebrook}(2014)}]{Obreschkow14b}
{Obreschkow} D., {Glazebrook} K., 2014, \apj, 784, 26

\bibitem[{{Okamoto} {et~al.}(2008){Okamoto}, {Gao}, \& {Theuns}}]{Okamoto08}
{Okamoto} T., {Gao} L., {Theuns} T., 2008, \mnras, 390, 920

\bibitem[{{Oosterloo} {et~al.}(2007){Oosterloo}, {Fraternali}, \&
  {Sancisi}}]{Oosterloo07}
{Oosterloo} T., {Fraternali} F., {Sancisi} R., 2007, \aj, 134, 1019

\bibitem[{{Padilla} {et~al.}(2014){Padilla}, {Salazar-Albornoz}, {Contreras},
  {Cora}, \& {Ruiz}}]{Padilla14}
{Padilla} N.~D., {Salazar-Albornoz} S., {Contreras} S., {Cora} S.~A., {Ruiz}
  A.~N., 2014, \mnras, 443, 2801

\bibitem[{{Pichon} {et~al.}(2011){Pichon}, {Pogosyan}, {Kimm}, {Slyz},
  {Devriendt}, \& {Dubois}}]{Pichon11}
{Pichon} C., {Pogosyan} D., {Kimm} T., {Slyz} A., {Devriendt} J., {Dubois} Y.,
  2011, \mnras, 418, 2493

\bibitem[{{Portinari} {et~al.}(1998){Portinari}, {Chiosi}, \&
  {Bressan}}]{Portinari98}
{Portinari} L., {Chiosi} C., {Bressan} A., 1998, \aap, 334, 505

\bibitem[{{Ruiz} {et~al.}(2013){Ruiz}, {Cora}, {Padilla}, {Dom{\'{\i}}nguez},
  {Tecce}, {Orsi}, {Yaryura}, {Garc{\'{\i}}a Lambas}, {Gargiulo}, \& {Mu{\~n}oz
  Arancibia}}]{Ruiz14}
{Ruiz} A.~N., {Cora} S.~A., {Padilla} N.~D., {Dom{\'{\i}}nguez} M.~J., {Tecce}
  T.~E., {Orsi} {\'A}., {Yaryura} Y.~C., {Garc{\'{\i}}a Lambas} D. {et~al},
  2013, ArXiv:1310.7034

\bibitem[{{Sales} {et~al.}(2012){Sales}, {Navarro}, {Theuns}, {Schaye},
  {White}, {Frenk}, {Crain}, \& {Dalla Vecchia}}]{Sales12}
{Sales} L.~V., {Navarro} J.~F., {Theuns} T., {Schaye} J., {White} S.~D.~M.,
  {Frenk} C.~S., {Crain} R.~A., {Dalla Vecchia} C., 2012, \mnras, 423, 1544

\bibitem[{{Schaye} {et~al.}(2014){Schaye}, {Crain}, {Bower}, {Furlong},
  {Schaller}, {Theuns}, {Dalla Vecchia}, {Frenk}, {McCarthy}, {Helly},
  {Jenkins}, {Rosas-Guevara}, {White}, {Baes}, {Booth}, {Camps}, {Navarro},
  {Qu}, {Rahmati}, {Sawala}, {Thomas}, \& {Trayford}}]{Schaye14}
{Schaye} J., {Crain} R.~A., {Bower} R.~G., {Furlong} M., {Schaller} M.,
  {Theuns} T., {Dalla Vecchia} C., {Frenk} C.~S. {et~al}, 2014, ArXiv e-prints

\bibitem[{{Serra} {et~al.}(2012){Serra}, {Oosterloo}, {Morganti}, {Alatalo},
  {Blitz}, {Bois}, {Bournaud}, {Bureau}, {Cappellari}, {Crocker}, {Davies},
  {Davis}, {de Zeeuw}, {Duc}, {Emsellem}, {Khochfar}, {Krajnovi{\'c}},
  {Kuntschner}, {Lablanche}, {McDermid}, {Naab}, {Sarzi}, {Scott}, {Trager},
  {Weijmans}, \& {Young}}]{Serra12}
{Serra} P., {Oosterloo} T., {Morganti} R., {Alatalo} K., {Blitz} L., {Bois} M.,
  {Bournaud} F., {Bureau} M. {et~al}, 2012, \mnras, 2823

\bibitem[{{Serra} {et~al.}(2014){Serra}, {Oser}, {Krajnovic}, {Naab},
  {Oosterloo}, {Morganti}, {Cappellari}, {Emsellem}, {Young}, {Blitz}, {Davis},
  {Duc}, {Hirschmann}, {Weijmans}, {Alatalo}, {Bayet}, {Bois}, {Bournaud},
  {Bureau}, {Davies}, {de Zeeuw}, {Khochfar}, {Kuntschner}, {Lablanche},
  {McDermid}, {Sarzi}, \& {Scott}}]{Serra14}
{Serra} P., {Oser} L., {Krajnovic} D., {Naab} T., {Oosterloo} T., {Morganti}
  R., {Cappellari} M., {Emsellem} E. {et~al}, 2014, ArXiv:1401.3180

\bibitem[{{Sharma} {et~al.}(2012){Sharma}, {Steinmetz}, \&
  {Bland-Hawthorn}}]{Sharma12}
{Sharma} S., {Steinmetz} M., {Bland-Hawthorn} J., 2012, \apj, 750, 107

\bibitem[{{Springel} {et~al.}(2005){Springel}, {White}, {Jenkins}, {Frenk},
  {Yoshida}, {Gao}, {Navarro}, {Thacker}, {Croton}, {Helly}, {Peacock}, {Cole},
  {Thomas}, {Couchman}, {Evrard}, {Colberg}, \& {Pearce}}]{Springel05}
{Springel} V., {White} S.~D.~M., {Jenkins} A., {Frenk} C.~S., {Yoshida} N.,
  {Gao} L., {Navarro} J., {Thacker} R. {et~al}, 2005, \nat, 435, 629

\bibitem[{{Sutherland} \& {Dopita}(1993)}]{Sutherland93}
{Sutherland} R.~S., {Dopita} M.~A., 1993, \apjs, 88, 253

\bibitem[{{Tenneti} {et~al.}(2014){Tenneti}, {Mandelbaum}, {Di Matteo}, {Feng},
  \& {Khandai}}]{Tenneti14}
{Tenneti} A., {Mandelbaum} R., {Di Matteo} T., {Feng} Y., {Khandai} N., 2014,
  \mnras, 441, 470

\bibitem[{{Vogelsberger} {et~al.}(2014){Vogelsberger}, {Genel}, {Springel},
  {Torrey}, {Sijacki}, {Xu}, {Snyder}, {Bird}, {Nelson}, \&
  {Hernquist}}]{Vogelsberger14}
{Vogelsberger} M., {Genel} S., {Springel} V., {Torrey} P., {Sijacki} D., {Xu}
  D., {Snyder} G., {Bird} S. {et~al}, 2014, \nat, 509, 177

\bibitem[{{Wardle} \& {Knapp}(1986)}]{Wardle86}
{Wardle} M., {Knapp} G.~R., 1986, \aj, 91, 23

\bibitem[{{Welch} {et~al.}(2010){Welch}, {Sage}, \& {Young}}]{Welch10}
{Welch} G.~A., {Sage} L.~J., {Young} L.~M., 2010, \apj, 725, 100

\bibitem[{{Wiklind} \& {Henkel}(1989)}]{Wiklind89}
{Wiklind} T., {Henkel} C., 1989, \aap, 225, 1

\bibitem[{{Young} {et~al.}(2011){Young}, {Bureau}, {Davis}, {Combes},
  {McDermid}, {Alatalo}, {Blitz}, {Bois}, {Bournaud}, {Cappellari}, {Davies},
  {de Zeeuw}, {Emsellem}, {Khochfar}, {Krajnovi{\'c}}, {Kuntschner},
  {Lablanche}, {Morganti}, {Naab}, {Oosterloo}, {Sarzi}, {Scott}, {Serra}, \&
  {Weijmans}}]{Young11}
{Young} L.~M., {Bureau} M., {Davis} T.~A., {Combes} F., {McDermid} R.~M.,
  {Alatalo} K., {Blitz} L., {Bois} M. {et~al}, 2011, \mnras, 414, 940

\bibitem[{{Young} {et~al.}(2013){Young}, {Scott}, {Serra}, {Alatalo}, {Bayet},
  {Blitz}, {Bois}, {Bournaud}, {Bureau}, {Crocker}, {Cappellari}, {Davies},
  {Davis}, {de Zeeuw}, {Duc}, {Emsellem}, {Khochfar}, {Krajnovic},
  {Kuntschner}, {McDermid}, {Morganti}, {Naab}, {Oosterloo}, {Sarzi}, \&
  {Weijmans}}]{Young13}
{Young} L.~M., {Scott} N., {Serra} P., {Alatalo} K., {Bayet} E., {Blitz} L.,
  {Bois} M., {Bournaud} F. {et~al}, 2013, ArXiv:1312.6318

\end{thebibliography}

\label{lastpage}
\appendix
\section[]{PDFs of angular momentum flips and slews}\label{PDFsP14}

\begin{figure}
\begin{center}
\includegraphics[trim = 1.5mm 0.5mm 1mm 1mm,clip,width=0.49\textwidth]{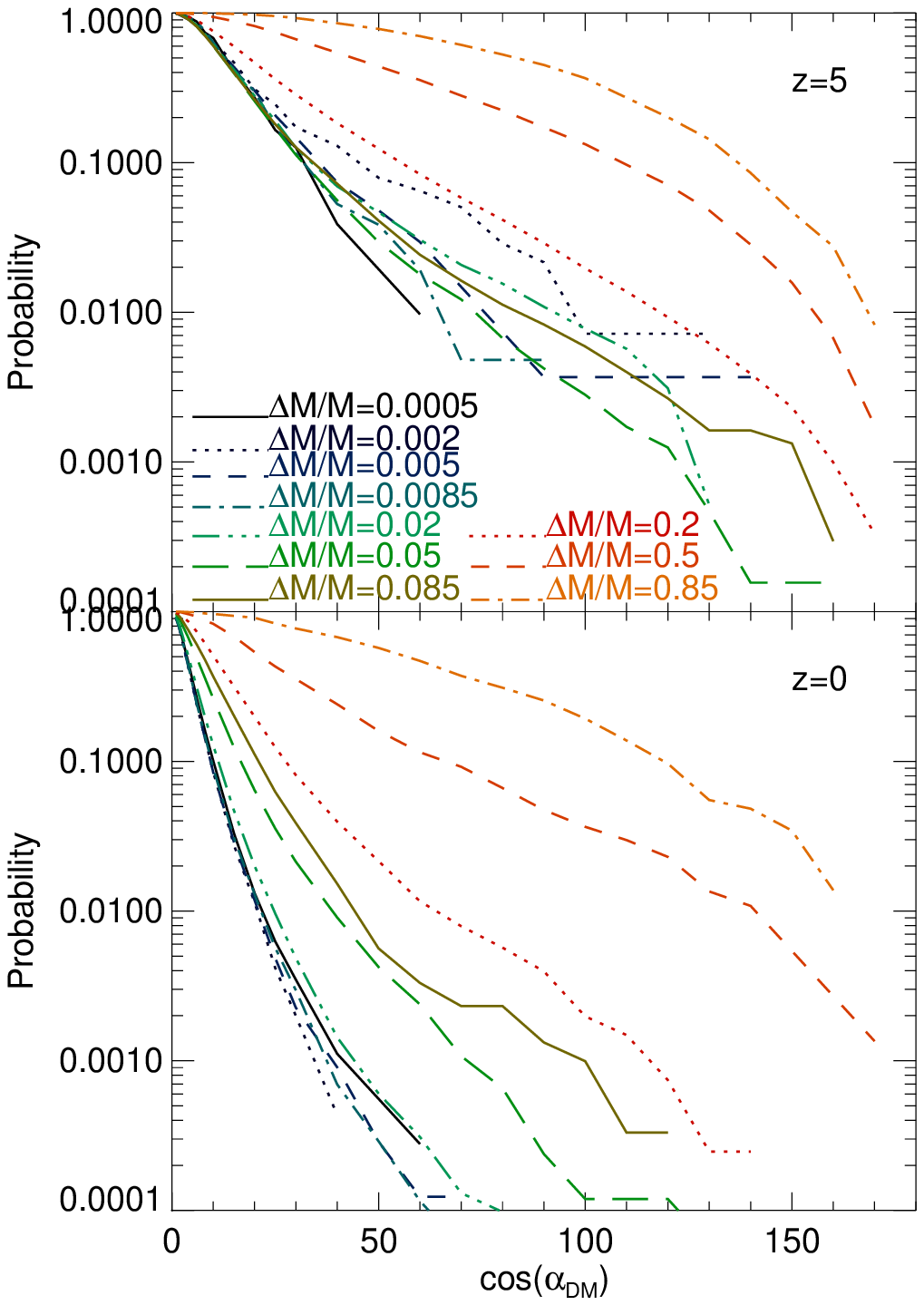}
\caption{PDFs of the angle between the angular momentum of dark matter halos before and after 
accretion of matter (excluding halo mergers). Each line corresponds to a different fractional change as labelled in the top panel.
The top panel shows PDFs at $z=5$, while the bottom panel shows the PDFs at $z=0$. These PDFs were calculated in the 
Millennium-II simulation by \citet{Padilla14}.}
\label{PDFs}
\end{center}
\end{figure}

\begin{figure}
\begin{center}
\includegraphics[trim = 1.5mm 0.5mm 1mm 1mm,clip,width=0.49\textwidth]{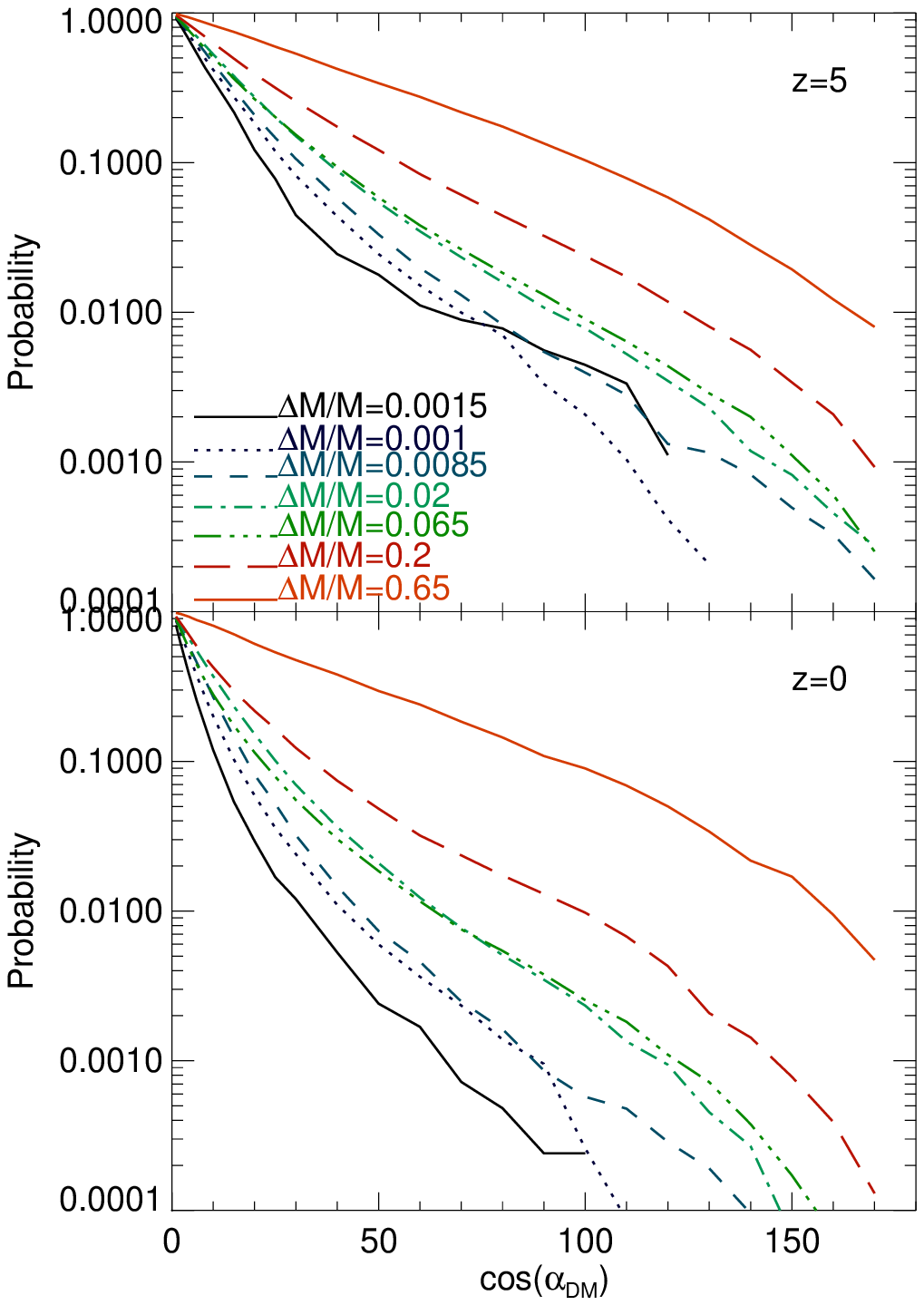}
\caption{As in Fig.~\ref{PDFs} but here we show the flips in the angular momentum of DM halos only due to mergers 
with other DM halos. The fractional change in mass due to mergers are as labelled in the top panel.}
\label{PDFs2}
\end{center}
\end{figure}

In Fig.~\ref{PDFs} we show the probability distribution functions 
for the angle between the angular momentum of dark matter halos in the Millennium-II 
before and after they have accreted mass as calculated by \citet{Padilla14}. 
Each line corresponds to a fractional change in mass and we show these PDFs for two different 
redshifts. Fig.~\ref{PDFs2} shows the same PDFs but for flips in the angular momentum due to 
mergers with other DM halos.

\end{document}